\documentclass[preprint,12pt]{elsarticle}

\usepackage{amssymb}

\usepackage{hyperref}

\journal{Ecological Modelling}
\biboptions{authoryear}

\begin{document}

\begin{frontmatter}

\title{Importance of spatial predictor variable selection in machine learning applications - Moving from data reproduction to spatial prediction}

\author[add1]{Hanna Meyer\corref{cor1}}
\ead{hanna.meyer@uni-muenster.de}
\cortext[cor1]{Corresponding author}
\author[add2]{Christoph Reudenbach}
\author[add2]{Stephan W{\"o}llauer}
\author[add2]{Thomas Nauss}

\address[add1]{Institute of Geoinformatics, Westf{\"a}lische Wilhelms-Universit{\"a}t M{\"u}nster, Heisenbergstr.2, 48149 M{\"u}nster}
\address[add2]{Faculty of Geography, Philipps-Universit{\"a}t Marburg, Deutschhausstr. 10, 35037 Marburg, Germany}

\frenchspacing

\begin{abstract}
Machine learning algorithms find frequent application in spatial prediction of biotic and abiotic environmental variables. However, the characteristics of spatial data, especially spatial autocorrelation, are widely ignored. We hypothesize that this is problematic and results in models that can reproduce training data but are unable to make spatial predictions beyond the locations of the training samples. We assume that not only spatial validation strategies but also spatial variable selection is essential for reliable spatial predictions.

We introduce two case studies that use remote sensing to predict land cover and the leaf area index for the ``Marburg Open Forest``, an open research and education site of Marburg University, Germany. We use the machine learning algorithm Random Forests to train models using non-spatial and spatial cross-validation strategies to understand how spatial variable selection affects the predictions.

Our findings confirm that spatial cross-validation is essential in preventing overoptimistic model performance. We further show that highly autocorrelated predictors (such as geolocation variables, e.g. latitude, longitude) can lead to considerable overfitting and result in models that can reproduce the training data but fail in making spatial predictions. The problem becomes apparent in the visual assessment of the spatial predictions that show clear artefacts that can be traced back to a misinterpretation of the spatially autocorrelated predictors by the algorithm. Spatial variable selection could automatically detect and remove such variables that lead to overfitting, resulting in reliable spatial prediction patterns and improved statistical spatial model performance.

We conclude that in addition to spatial validation, a spatial variable selection must be considered in spatial predictions of ecological data to produce reliable predictions.

\end{abstract}


\end{frontmatter}


\section{Introduction}

A key task in ecology is studying the spatial or spatio-temporal patterns of ecosystem variables, e.g. climate dynamics \citep{Appelhans2015}, variability of soil properties \citep{Gasch2015} or distribution of vegetation types \citep{Juel2015}. Spatially continuous datasets of ecosystem variables are needed to analyze the spatial patterns and dynamics. However, ecological variables are typically acquired through field work, which only provides data with a limited spatial extent, such as from climate stations, soil profiles or plot-based vegetation records. These data do not provide spatially continuous information about the variable of interest. 
Predictive modelling is a method commonly used to derive spatially continuous datasets from limited field data \citep[e.g.][]{Lary2016}. In predictive modelling, field data is used to train statistical models using spatially continuous predictor variables derived from remote sensing imagery. The resulting model is then used to make predictions in space, i.e. beyond the locations used for model training. 

Most contemporary predictive modelling approaches use flexible machine learning algorithms, which can approximate the nonlinear and complex relationships found in nature.
Recent software developments have simplified the application of machine learning algorithms \citep[e.g. for R see][] {Kuhn2013}. Noteworthy, however, is that machine learning is applied to ecological spatial modelling the same way as it is in other disciplines, while ignoring the unique characteristics of spatial environmental data. Yet, spatial (and temporal) dependencies differentiate spatial data from ``ordinary`` data and complicate the use of machine learning – due to the nature of the data, we cannot assume samples are identically and independently distributed (i.i.d assumption) \citep{Xie2017}. This is especially true when data are sampled in clusters, which is a common design for providing ground truth data used in predictive modelling of ecological data.

Previous studies in spatial applications of machine learning algorithms have widely ignored the spatial dependencies in the data.
One problem of ignoring spatial dependencies in prediction methods becomes obvious in the error assessment of spatial predictive models. Many authors have shown that the commonly used random cross-validation provides considerably overoptimistic error estimates due to the problem of autocorrelation \citep{Bahn2013, Micheletti2014,Juel2015,Gasch2015,Gudmundsson2015,Roberts2017,Meyer2018}. Hence cross-validation strategies based on random data splitting fail to assess a model’s performance in terms of spatial mapping and only validate its ability to reproduce the sampling data.
Several methods for spatial cross-validation have been proposed to account for spatial dependencies in the data \citep{Brenning2005,LeRest2014, Pohjankukka2017, Roberts2017, Meyer2018, Valavi2018}. While spatial cross-validation can solve this issue and provide objective and meaningful error estimates, the algorithms' strong performance with random subsets and complete failures when predictions are made beyond the spatial extent of the training samples still remains an issue.

\citet{Meyer2018} have shown for spatio-temporal data that spatial (or spatio-temporal) dependencies cause a misinterpretation of certain predictor variables which makes flexible algorithms fail when predicting beyond the location of the training data. Spatial dependencies in predictor variables are most apparent in ``geolocation`` predictors that describe the spatial location of the training samples (e.g. coordinates, elevation, euclidean distances and all derivations of these data). Hence, we assume that including predictor variables that describe the spatial location are problematic and prevent spatial models from making meaningful contributions to ecological research. However, predictor variables that describe the spatial location rather than the environmental properties are commonly included. Spatial coordinates are used especially often \citep{Li2011,Langella2010, Shi2015a,Janatian2017, Walsh2017, Jing2016, Wang2017, Georganos2019}. Distances to certain points \citep[e.g.][]{Hengl2018} or Euclidean distance to the corner coordinates of the model domain \citep[e.g.][]{Behrens2018} have also been suggested as predictors and included in models.

This study uses autocorrelated spatial data to investigate the sensitivity of machine learning applications to commonly applied geolocation predictors and shows pathways towards an automatic selection of predictors that cannot be incorporated in spatial prediction tasks.
We assume that spatial models cannot handle predictor variables that are highly autocorrelated in space (e.g. geolocation) due to spatial dependencies in the training data. Algorithms can easily misinterpret such variables, leading the model to make erroneous predictions outside of the locations of the training data. The problem becomes obvious in the limited spatial performance of the model as well as in visually obvious artefacts in the spatial predictions.
We therefore assume that spatial variable selection is essential for automatically removing variables counterproductive to spatial mapping to provide scientifically valuable results.

We use two examples of classic prediction tasks in environmental science to investigate our hypotheses. First, we perform a Land Use/ Land Cover (LULC) classification, which is a common field for applying machine learning-based predictive modelling in the context of ecology and remote sensing. The study area is located around the ``Marburg Open Forest``, an open research and education site owned by Marburg University in Hessen, Germany. Second, we model the Leaf Area Index (LAI) for the same region. Spectral, terrain-related as well as geolocation variables are used as potential predictor variables in both examples.
We study the effect of spatial and non-spatial cross-validation on the estimated model performance with the frequently applied machine learning algorithm Random Forest. A spatial variable selection is suggested to analyze the importance of the potential predictor variables for spatial mapping and their effect on the prediction outcomes.

\section{Methods}

The following sections describe the two case studies, the data, all processing steps, modelling as well as validation.
Data processing and modelling were performed in R Version 3.4 \citep{RCT2018}.
The scripts performing processing and analysis can be retrieved from \url{https://github.com/HannaMeyer/EcoMod_SpML}.

\subsection{Prediction task I: Land use/land cover classification}
\begin{figure}[htbp]
\includegraphics[width=1\textwidth]{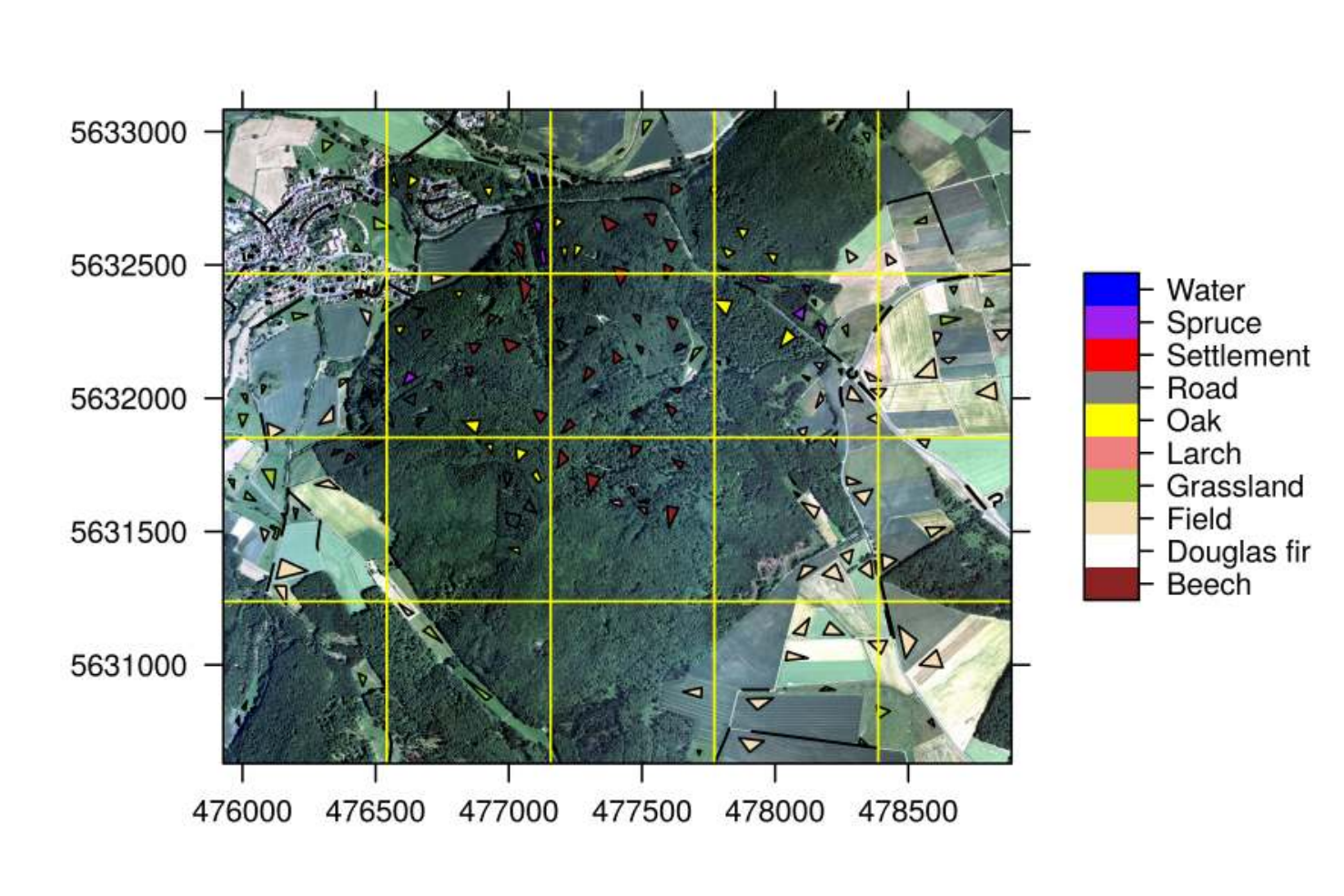}
\caption{Study area represented by the true color composite of the aerial image to be classified. Polygons indicate the training areas of the different LULC classes used for model training. The yellow grid represents spatial blocks used for spatial cross-validation.
Reference system: UTM 32N (WGS84).}
\label{fig:casestudy_LUC} 
\end{figure}

The first prediction task is to classify different types of forest, as well as adjacent LULC for the ``Marburg Open Forest`` \url{http://nature40.org} in Hessen, Germany.
The basis for the classification is an aerial image that covers approx. 3000 x 2500~m (Fig.~\ref{fig:casestudy_LUC}).

\subsubsection{Reference data}
A set of manually digitized polygons covering typical LULC classes are used as reference data which were selected by a combination of visual image inspection and knowledge firsthand from field work. In total, 10 different LULC classes were assigned (Table~\ref{tab:dataset}).

\begin{table}[htb]
\small
\centering
\caption{Summary of the different land use/land cover classes and the size of training data used for training of the classification model.} 
\label{tab:dataset}
\begin{tabular}{lcc}
\hline
Type & Polygons & Pixels  \\ \hline

Beech & 34  & 31306  \\
Douglas fir& 20  & 13241  \\
Field& 40  & 59663  \\
Grassland& 85  & 27134  \\
Larch& 4  & 1568  \\
Oak& 23  & 17804  \\
Road& 38  & 18461  \\
Settlement& 40  & 4722  \\
Spruce& 14  & 7521  \\
Water& 66  & 3261 \\ \hline

\end{tabular}
\end{table}

\subsubsection{Predictor variables}
\label{sec:LUC_predictors}

\begin{figure}[htbp]
\includegraphics[width=1\textwidth]{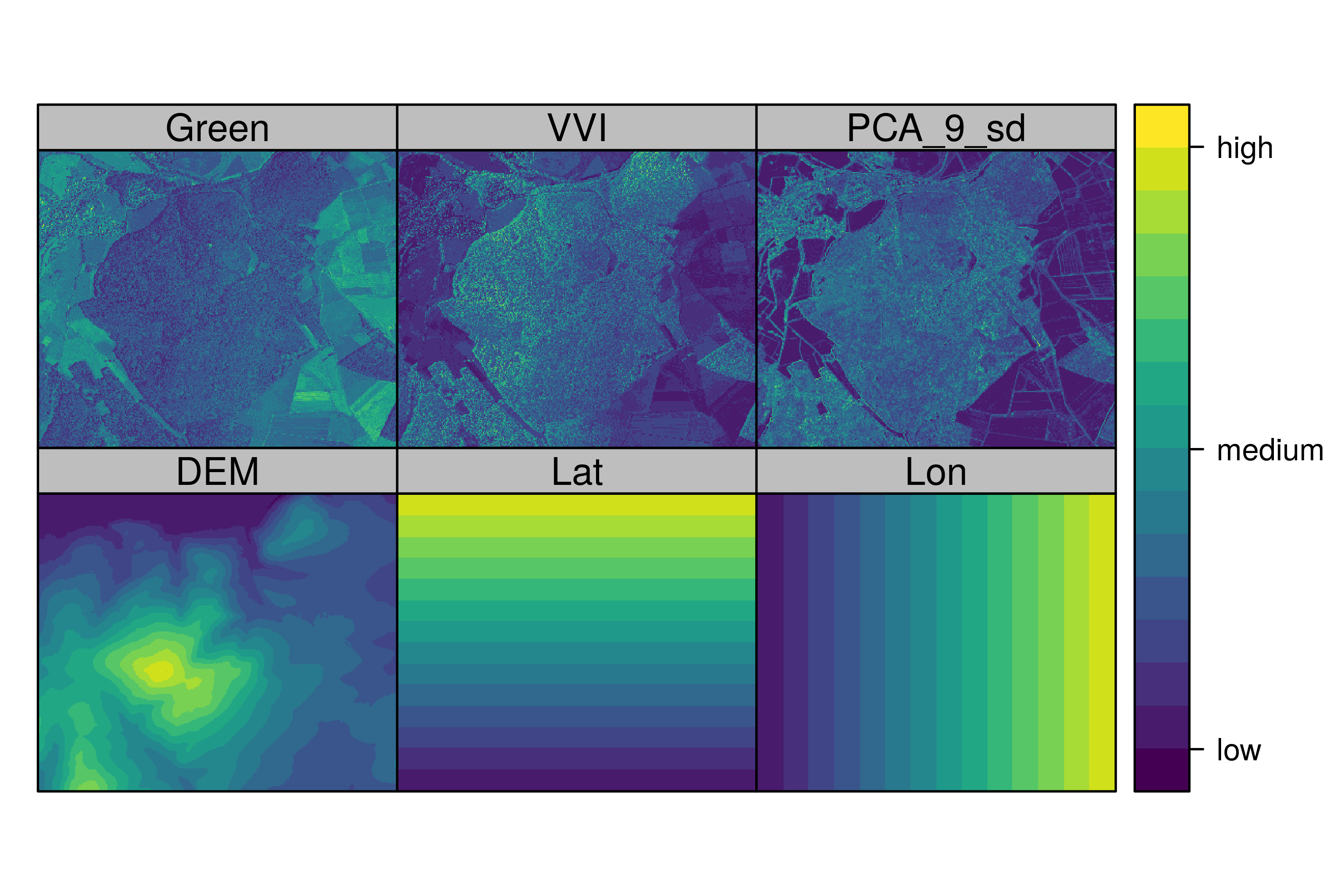}
\caption{Example of spectral, terrain-related and geolocation predictor variables: reflectance in the green band (green), Visible Vegetation Index (VVI), standard deviation in a 9x9 pixel environment of the first Principal Component of all spectral variables (PCA\_9\_sd), Digital Elevation Model (DEM), latitude (Lat) and longitude (Lon).}
\label{fig:predictors} 
\end{figure}

Spectral, terrain-related and geolocation variables were all prepared as potential predictor variables (Fig.~\ref{fig:predictors}).
Spectral variables come from a 20~cm resolution aerial image \citep{HVBG2018a} taken at the 30th of September 2015. For this study, the image was resampled to a spatial resolution of 1m.
The spectral predictors were the three channels of the aerial image (red, green, blue). Further, the Visible Vegetation Index \citep[VVI,][]{PHL2015}, Triangular Greenness Index \citep[TGI,][]{Hunt2013}, Normalized Green Red Difference Index (NGRDI), and Green Leaf Index \citep[GLI,][]{Hunt2013}) were derived from the channels. A Principal Component Analysis (PCA) was performed on the three channels of the visible spectrum and the vegetation indices; the first component of the PCA was included as an additional potential predictor variable. In addition, the standard deviation of the first principal component was calculated in 3x3 (PCA\_3\_sd), 5x5 (PCA\_5\_sd) and 9x9 (PCA\_9\_sd) pixel environments to account for the spectral variability of LULC classes. A lidar-derived 1m Digital Elevation Model (DEM) was used as a terrain-related predictor. Elevation in the study area ranges from 210 to 415~m. Slope and aspect were calculated from the DEM in radians. 
The geolocation variables considered as potential predictors were latitude (Lat) and longitude (Lon).
This results in a total set of 16 predictor variables for LULC prediction.

\subsubsection{Compilation of the training data set}

The set of predictor variables was extracted for each training polygon. The model considered each pixel related to the polygons (e.g. within, intersecting) as an individual training sample (Table~ \ref{tab:dataset}). This resulted in a set of approx. 185000 training samples. Each training sample contained the information about every potential predictor variable as well as about the LULC class based on the information from the polygons.

\subsection{Prediction task II: Leaf Area Index modelling}
The second prediction task aimed at modelling the LAI for the forested area of the Marburg Open Forest, a classic example of a regression task in environmental science.

\subsubsection{Reference data}

\begin{figure}[htbp]
\includegraphics[width=1\textwidth]{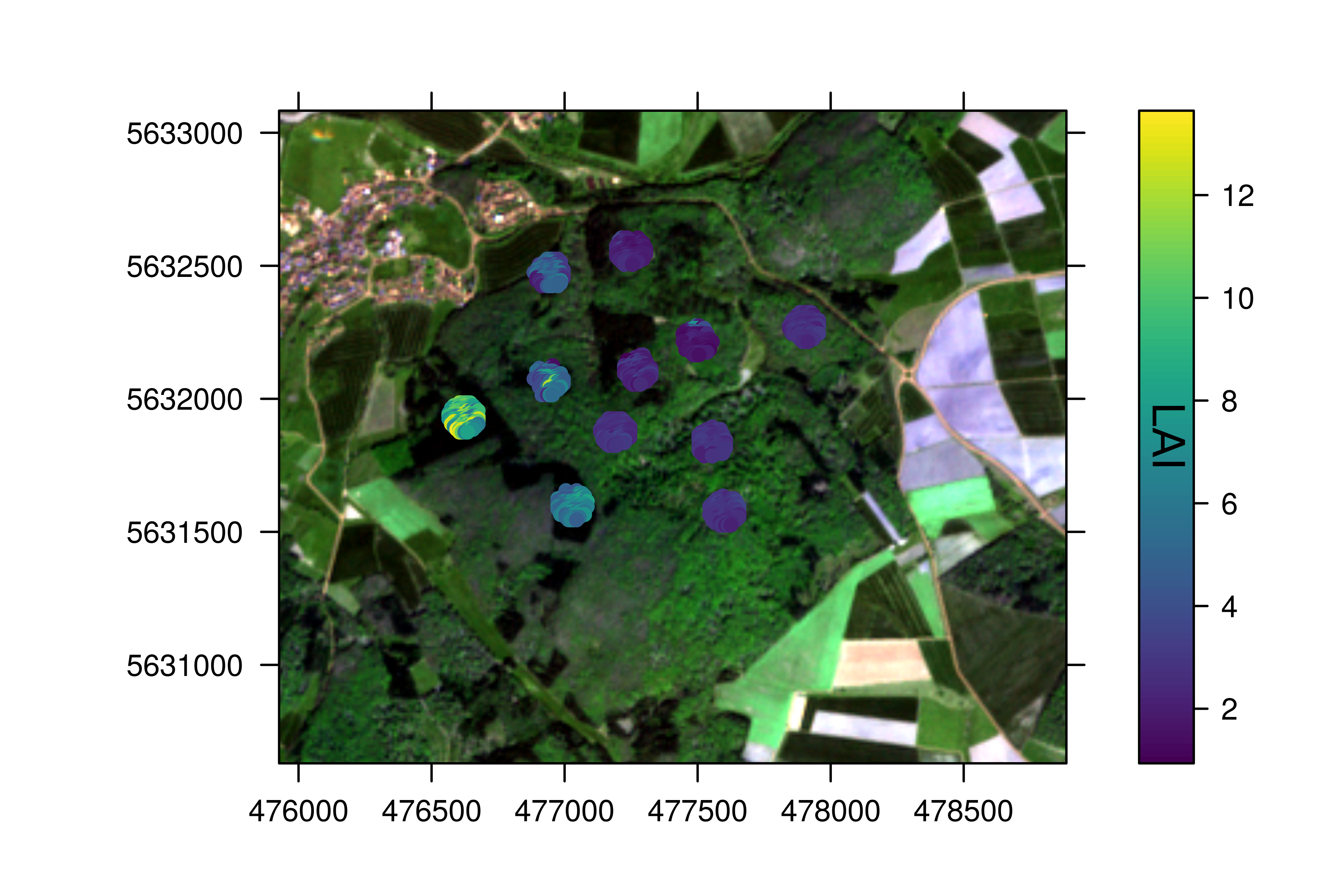}
\caption{Study area represented by the true color composite of the Sentinel-2 scene, which serves as the baseline for the LAI predictions. Points represent the location of the training samples. The color indicates the LAI values at these locations as derived from the lidar. The clear spatial clusters are the baseline for the spatial cross-validation.
Reference system: UTM 32N (WGS84).}
\label{fig:casestudy_LAI} 
\end{figure}

In this case study, the LAI reference was derived from lidar data taken in the vegetation period 2010 \citep{HVBG2018} that have 15cm vertical and 30cm spatial accuracy. The LAI was calculated from the lidar point cloud according to \citet{Getzin2017}. Since no major management was present in the forest, the LAI from the lidar data was regarded as a reference for this study despite the time lag between lidar derived reference and the Sentinel-2 based predictor variables. Especially since this study focus on the effect of validation and variable selection strategies this time lag was neglected. 
The calculated LAI data was then rasterized with 10m spatial resolution to match the geometry of the Sentinel-2 data that were used as predictors. To do this, the mean of all LAI values located in the extent of a Sentinel-2 pixel was calculated.
11 spatially distinct clusters were then assigned in homogenous areas of the forest. Every pixel in a 60m radius around the center of each cluster was used as training data, resulting in clear spatial clusters of training samples (Fig.~\ref{fig:casestudy_LAI}). In total, 824 training pixels distributed across the 11 clusters were used. The minimum, maximum and mean LAI in this training data set were 0.9, 13.6 and 4.2, respectively. A LAI below 1 means that the area is not fully covered by leaves. Values larger than 1 mean that more than one layer of leaves are present.

\subsubsection{Predictor variables}
\label{sec:LAI_predictors}
A Sentinel-2 scene as Level-1C product from 2017/05/10 was used to derive spectral predictor variables. Sentinel-2 is the  optical system from the earth observation mission from the EU Copernicus Programme and has channels in the visible (bands 2-4), red edge (bands 5-7), near infrared (band 8 and 8A) and short-wave infrared (bands 11 and 12) part of the spectrum. The Sentinel-2 bands 1, 9 and 10 were not considered in this study because they don't include relevant information for this prediction task. Hereafter, the used channels are referred to as B01, B02,...B12. Channels B02-B04 and B08 have a spatial resolution of 10m. The other channels have a resolution of 20~m and were resampled to match the geometry of the 10~m channels. In addition to the spectral channels, elevation, slope, aspect, as well as latitude and longitude (as described in the description of the previous prediction task but resampled to a 10m spatial resolution) were used as potential predictors. This results in a total set of 15 predictor variables for LAI prediction.

\subsubsection{Compilation of the training data set}

Values for each predictor variable was extracted from the locations of the training samples. Each training sample contained the extracted information from all potential predictor variables as well as the information about the LAI based on the information from the lidar-derived reference points.

\subsection{Model training and prediction}

The Random Forest algorithm \citep{Breiman2001} was chosen as the machine learning algorithm for predictive modelling due to its prominence in ecological modelling.
Random Forest bases on the concept of regression and classification trees: a series of nested decision rules for the predictors determine the response (also called reference, i.e. LULC or LAI). Random forest repeatedly builds trees from random samples of the training data. Each tree is a separate model of the ensemble. The predictions of all trees are averaged to produce the final estimate. To overcome correlation between trees, a number of predictors (mtry) are randomly selected at each split. The best predictor from the random subset is used at this split to partition the data.

In this study, the Random Forests implementation of the \textsf{randomForest} package \citep{Liaw2002} in R was applied and accessed via the \textsf{caret} package \citep{Kuhn2016}. Throughout the study, each Random Forest model consisted of 500 trees after no increase in performance could be observed using a higher number of trees. The number of randomly chosen predictor variables at each split of the tree (``mtry``) was tuned between two and the number of predictor variables (16 for LULC predictions and 15 for LAI predictions). See \citet{Kuhn2013} for a more detailed description on the Random Forest algorithm and mtry tuning.

\begin{figure}[htbp]
\includegraphics[width=1\textwidth]{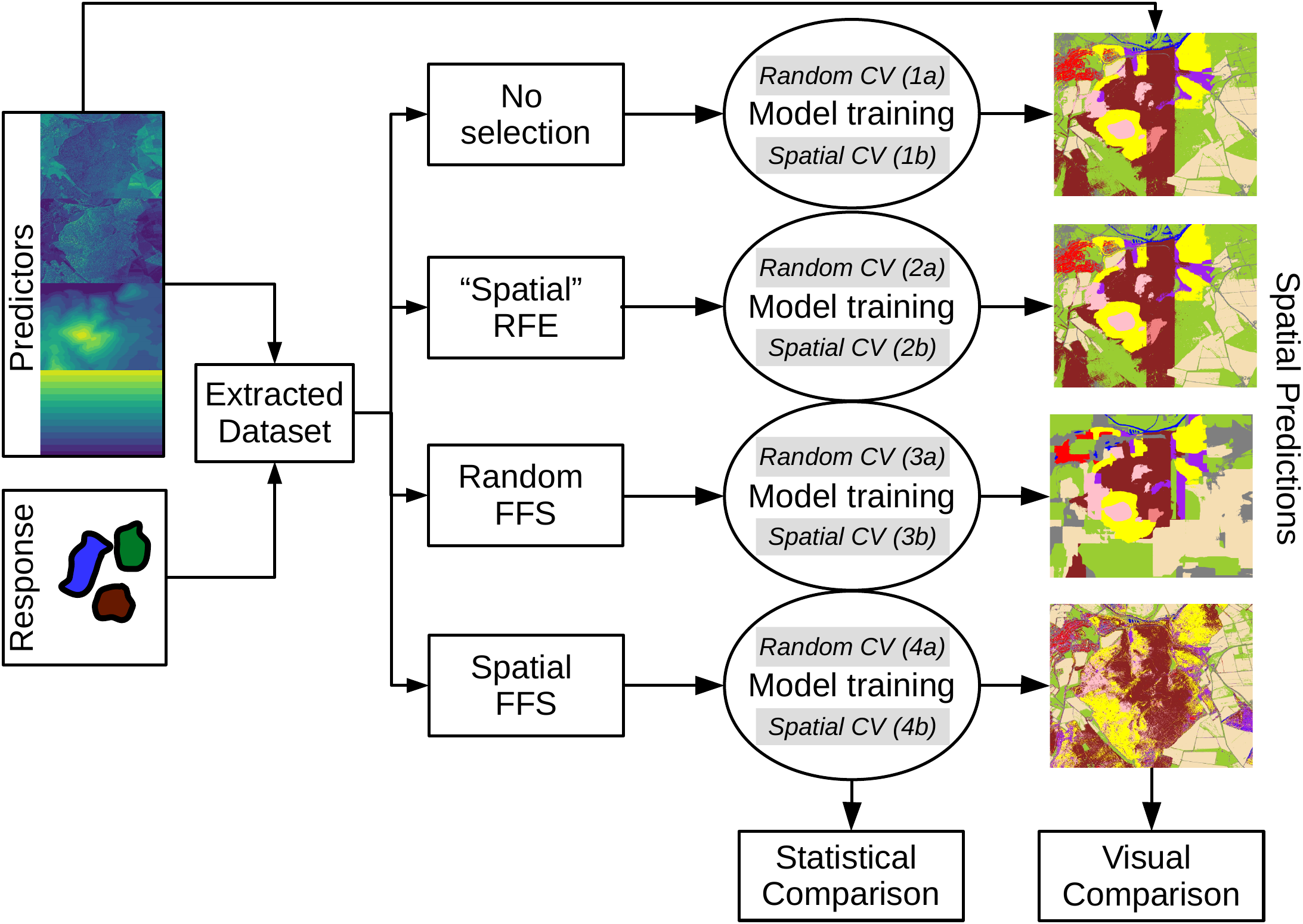}
\caption{Overview on the models compared in this study. Spectral, terrain related as well as geolocation variables were used as predictors (only examples shown here). The response variable was derived from training polygons for the case study of land cover classification or from lidar-derived Leaf Area Index (LAI) values for the case study of LAI prediction. Models were trained using either no variable selection or either recursive feature elimination, default random forward feature selection (FFS) or spatial FFS. Model performance was compared with random cross-validation (CV) or spatial CV. The entire modelling procedures were performed for the case study of land cover classification as well as for prediction of LAI.}
\label{fig:flowchart} 
\end{figure}

To study the effect of spatial validation as well as spatial variable selection the following models were compared for both case studies:

\begin{enumerate}{
\item Model using \textbf{all} potential predictor variables. Performance was estimated by \textbf{random} cross-validation (see 1a in Fig.~\ref{fig:flowchart}). The results of this model are used to show the outcome of a ``default`` modelling approach. The performance was further estimated by \textbf{spatial} cross-validation (see 1b in Fig.~\ref{fig:flowchart}). The results of this validation are used to show how spatial cross-validation affects the estimated error of a ``default`` model.

\item Model using \textbf{selected} variables only. Variable selection was based on the commonly used recursive feature elimination with \textbf{spatial} cross-validation. Performance was estimated by \textbf{random} cross-validation (see 2a in Fig.~\ref{fig:flowchart}) and \textbf{spatial} cross-validation (see 2b in Fig.~\ref{fig:flowchart}). The results of this model are used to show how "default" variable selection affects the spatial model performance.

\item Model using \textbf{selected} variables only. Variable selection was based on a forward feature selection with \textbf{random} cross-validation. Performance was estimated by \textbf{random} cross-validation (see 3a in Fig.~\ref{fig:flowchart}) and \textbf{spatial} cross-validation (see 3b in Fig.~\ref{fig:flowchart}). The results of this model are used to show how spatial variable selection affects the random model performance.

\item Model using \textbf{selected} variables only. Variable selection was based on a forward feature selection with \textbf{spatial} cross-validation. Performance was estimated by \textbf{random} cross-validation (see 4a in Fig.~\ref{fig:flowchart}) and \textbf{spatial} cross-validation (see 4b in Fig.~\ref{fig:flowchart}). The results of this model are used to show how spatial variable selection affects the spatial model performance.
}
\end{enumerate} 

The following sections describe the different cross-validation and variable selection strategies in more detail.

\subsubsection{Cross-validation strategies}

\begin{figure}[htbp]
\includegraphics[width=1\textwidth]{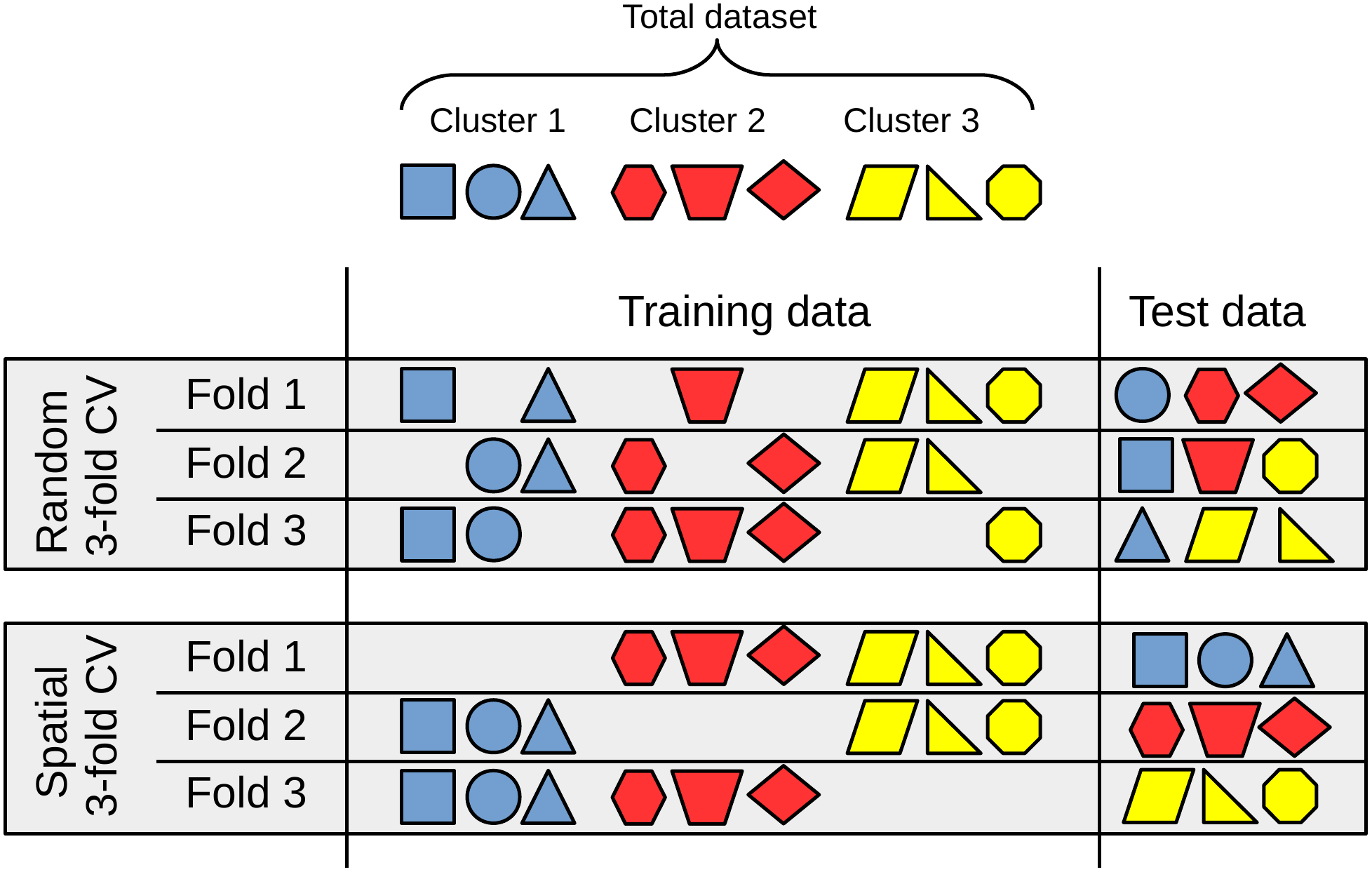}
\caption{Concept of random and spatial cross-validation (CV): A total dataset (here: 9 different data points represented by different shapes) is split into k folds (here: k=3). Models are then repeatedly trained by always leaving one of the folds out and use it for model validation and not for model training. Random CV means that the data are randomly split into folds. Spatial CV means that the data are split into folds according to spatial location (e.g. a spatial cluster or a spatial block, here represented by unique color). Figure modified from \citet{Meyer2018} and \citet{Kuhn2013}.}
\label{fig:cv_concept} 
\end{figure}

This study applied two cross-validation strategies: a standard random k-fold cross-validation and a spatial k-fold cross-validation. Each strategy first splits the data into k folds and then repeatedly trains the models (k times) using the data from all but one fold. The models are evaluated based on how they perform with the left-out data (see Fig.~\ref{fig:cv_concept}). See \citep{Kuhn2013} for more detailed description on cross-validation in general.

While the cross-validation procedure is the same for random and spatial cross-validation, the major difference is how the data points are split into folds (see Fig.~\ref{fig:cv_concept}). 
For the standard random cross-validation, each data point was \textbf{randomly} assigned to one of the k folds. 

For the spatial cross-validation, we chose a spatial block approach as suggested in \citet{Roberts2017} and also \citet{Valavi2018} for the LULC case study. Therefore, we divided the spatial domain into 20 equally sized spatial blocks (yellow grid in Fig.~\ref{fig:casestudy_LUC}). For each training sample, the spatial block affiliation was identified by the spatial location of the corresponding training polygon where the sample belongs to. If a training polygon lay within two spatial blocks, the sample was only assigned to the one block in which the greater proportion of the polygon lay. This precluded that training pixels from one (usually homogeneous) polygon were present in two spatial blocks.
Analogous to the random cross-validation, models were then repeatedly trained using data from all but one spatial block (= fold) and their performance estimated using the left-out data, e.g. the spatial block left out of model training. Hence, models were assessed for their performance in making spatial predictions beyond the locations of the training data.
For the case study of the LAI predictions, a leave-one-cluster-out cross-validation was applied. This method is similar to the concept described above but instead of spatial blocks, in each iteration one of the 11 clusters Fig.~\ref{fig:casestudy_LAI}) was left out during model training.

The number of spatial blocks or clusters, k, equaled the number of spatial blocks (20) or the number of spatial clusters (11), depending on the case study. Random cross-validation was performed with the same number for k.

The validation measure for performance assessment for the LULC classification during cross-validation was the Kappa Index \citep{Cohen1960} and the Accuracy. A Kappa of 0 (or lower) is associated with a random classification result, while a Kappa of 1 indicates a perfect classification. Since it accounts for chance agreement, it was used as the prior validation measure for the classification task rather than the Accuracy. For the LAI predictions, the performance was assessed by the Root-Mean-Square Error (RMSE) and the coefficient of determination ($R^2$). 

Models were compared by fold, which gives the average performance (e.g. Kappa) over all k folds from cross-validation. Models were also compared by their global performance, which results from a comparison between every data point predicted during cross-validation, independently of the fold. 

\subsubsection{Spatial predictor variable selection}

Forward Feature Selection (FFS) as described in \citet{Meyer2018} and implemented in the CAST package \citep{Meyer2018c} for R was used to test the effect of spatial variable selection. This FFS implementation works in conjunction with user-defined cross-validation, hence it allows to select variables that lead to the highest spatial performance (if run in conjunction with spatial CV). First, the FFS trains models using every combination of two-predictor variables. Based on the best performing variables (as identified by cross-validation), FFS increases the number of variables to test which (if any) variables further improve the (spatial) model’s performance. Every variable that does not improve (or even decreases) the model’s performance is excluded. See \citet{Meyer2018} for a more detailed description on this FFS.

This study used FFS with spatial cross-validation to test which variables are significant to spatial mapping and which ones have no spatial meaning or are even counterproductive (spatial FFS, Fig.~\ref{fig:flowchart}). For comparison, FFS was also run with random cross-validation (random FFS, Fig.~\ref{fig:flowchart}) to check that improvements were indeed due to the spatial selection and not the pure reduction of variables that lead to changes in performance.
As FFS is very time consuming, mtry was set to 2 for feature selection. Once the variables were selected, models were re-trained using the selected variables and either spatial or random cross-validation with mtry being tuned between 2 and the number of selected predictor variables.

In addition to FFS, we also used recursive feature elimination \citep[RFE, explained in ][]{Kuhn2013} to compare state-of-the art procedures \citep[see e.g.\@][in the field of environmental mapping]{Brungard2015,Meyer2017a, Meyer2017, Ghosh2014, Stevens2013}. However, we argue that the backward RFE selection fails to address the issue of overfitting. RFE relies on variable importance scores, which are only calculated using the training subset \citep{Kuhn2013}.
If a variable leads to considerable overfitting, it is highly significant in the models. Therefore, the RFE process will mark it as important and not remove it, even if it results in a high spatial error. A forward selection in conjunction with spatial CV is therefore required.

\section{Results} 

\subsection{Statistical performance}
\begin{figure}[htbp]
\includegraphics[width=1\textwidth]{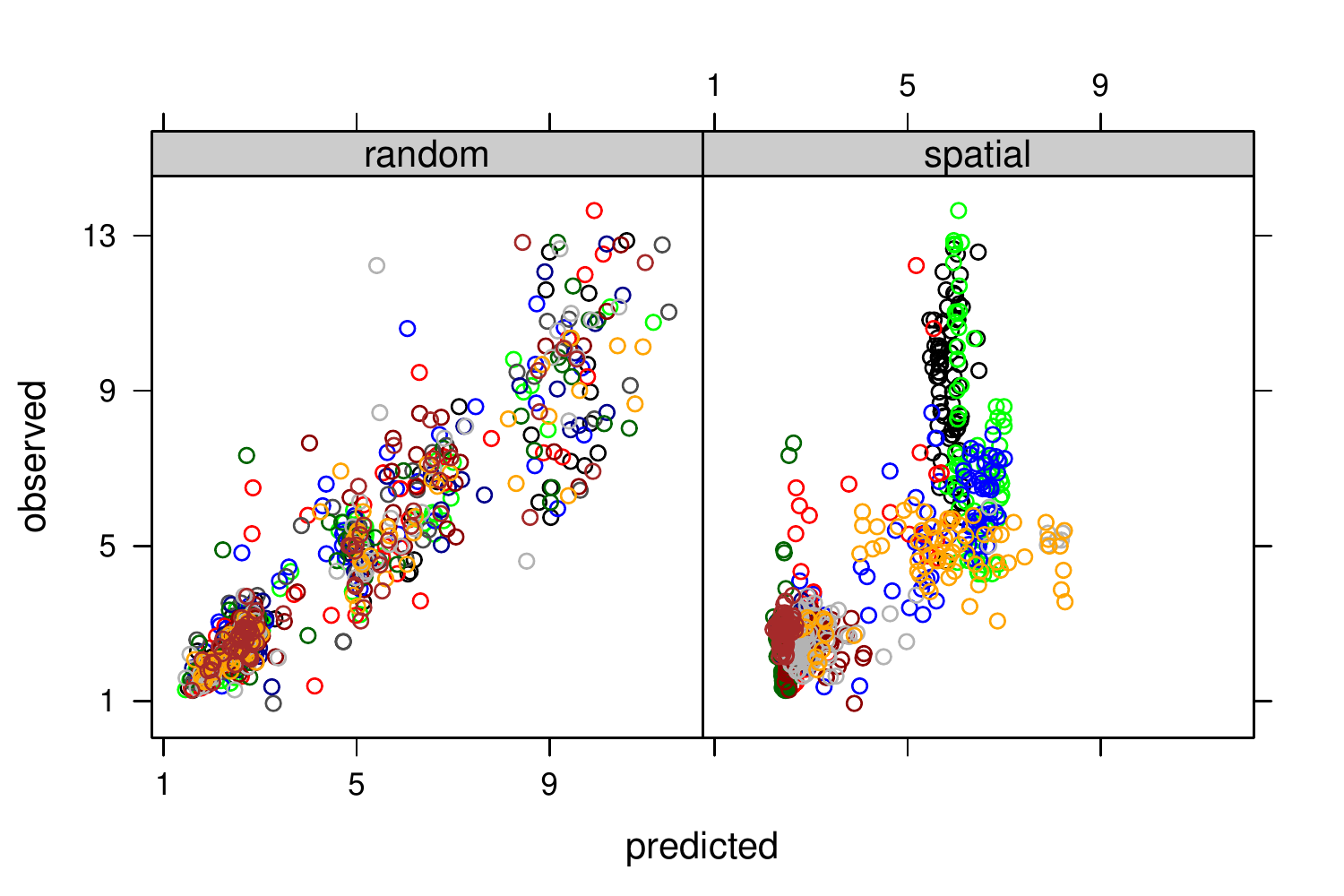}
\caption{Comparison between observed and predicted LAI values based on random (a) and spatial (b) cross-validation. Colors indicate the individual 11 folds. 
Note that the resulting models are quasi identical regardless of the validation strategy being used since cross-validation in this case serves mtry tuning and validation purposes only.}
\label{fig:perf_scatter} 
\end{figure}

Using the ``default`` way of spatial prediction (using all potential variables) and the ``default`` random cross-validation, both Accuracy and Kappa index were higher than 0.99 for the classification task (Table~\ref{tab:validation_LUC}). Random cross-validation indicated that the LAI could be predicted with a RMSE of 0.96 and a $R^2$ of 0.87 (Table~\ref{tab:validation_LAI}, Fig.~\ref{fig:perf_scatter}a).

When these models were validated using a spatial cross-validation, the performance was considerably lower (Kappa value of 0.55 for the LULC classification and RMSE of 1.25 for the LAI regression model, Table~\ref{tab:validation_LAI}, Fig.~\ref{fig:perf_scatter}b). A prominent source of high error estimates is that by leaving entire clusters out for validation, a held back cluster that has higher LAI values than all other clusters could not adequately be modelled since such high LAI values are unknown from the training data (Fig.~\ref{fig:perf_scatter}b). Note that low per-fold $R^2$ values for the LAI regression models in Table~\ref{tab:validation_LAI} result from low variabilities within spatial folds (Fig.~\ref{fig:perf_scatter}b) so that the per-fold RMSE or especially the global $R^2$/RMSE present the more reliable performance estimates here. 

Using an RFE-based variable selection in conjunction with a spatial cross-validation during the variable selection does not (LULC classification Kappa = 0.55) or only marginally (LAI regression RMSE = 1.22) improve the spatial performances. The same is true for an FFS-based approach in conjunction with a random cross-validation during the variable selection (Kappa = 0.14, RMSE = 1.23). In both cases, the random model performance stayed high (Kappa $>$ 0.99 for both RFE and random FFS, RMSE = 0.93 for RFE and 0.91 for FFS random). Hence, neither a FFS with random selection nor the RFE approach does prevent spatial overfitting even though spatial folds have been used for the latter.

When FFS was paired with spatial cross-validation (spatial FFS) for the variable selection task, the spatial performance slightly improved (Kappa = 0.56, RMSE = 1.20) compared to all other models. It is noteworthy that this type of variable selection reduces the model performance indicators in a random cross-validation (Kappa=0.87, RMSE= 1.12) so that the differences in the error estimates between the validation strategies became smaller. This validation was based on the average performance for each fold that was left out during cross-validation. Similar patterns emerged when all independent predictions were simultaneously compared (global validation). Noticeable is that the $R^2$ of LAI predictions increased from 0.58 (all variables, spatial cross-validation) to 0.63 (spatial variable selection, spatial cross-validation).

\begin{table}[htbp]
\small
\centering
\caption{Statistical performance of the models for LULC classification. Models were compared by fold, which gives the average performance over all k folds from cross-validation (CV). Models were also compared by their global performance, which is the Accuracy or Kappa for every data point predicted during cross-validation. Bold numbers indicate a spatial validation that must be considered as the valid performance for the prediction task. For an overview on the model-ID see also Fig.~\ref{fig:flowchart}.}
\label{tab:validation_LUC}
\begin{tabular}{lllllll}
\hline
&&&\multicolumn{2}{c}{By Fold} & \multicolumn{2}{c}{Global} \\
\hline
ID &Variables & CV & Accuracy & Kappa &  Accuracy &Kappa\\ \hline

1a &all & random  & $>$0.99 & $>$0.99 & $>$0.99 & $>$0.99 \\
1b &all & spatial  & \textbf{0.71} & \textbf{0.55}& \textbf{0.68} & \textbf{0.61} \\
2a &selected by RFE ``spatial`` & random & $>$0.99 & $>$0.99 &  $>$0.99 & $>$0.99\\
2b &selected by RFE ``spatial`` & spatial &\textbf{0.71} & \textbf{0.55} &  \textbf{0.69} &\textbf{0.61}\\
3a &selected by FFS random & random  & $>$0.99 & $>$0.99 & $>$0.99 & $>$0.99  \\
3b &selected by FFS random & spatial  &\textbf{ 0.43}  & \textbf{0.14} & \textbf{0.41} & \textbf{0.30}\\
4a&selected by FFS spatial & random  & 0.89 &  0.87 &  0.78 & 0.82\\
4b&selected by FFS spatial & spatial  & \textbf{0.71} & \textbf{0.56} & \textbf{0.70} & \textbf{0.62}\\
\hline
\end{tabular}
\end{table}

\begin{table}[htbp]
\small
\centering
\caption{Statistical performance of the models for LAI prediction. Models were compared by fold, which gives the average performance over all k folds from cross-validation (CV). Models were also compared by their global performance which is the RMSE or $R^2$ over every data point predicted during cross-validation. Bold numbers indicate a spatial validation that must be considered as the valid performance for the prediction task. For an overview on the model-ID see also Fig.~\ref{fig:flowchart}.}
\label{tab:validation_LAI}
\begin{tabular}{lllllll}
\hline
&&&\multicolumn{2}{c}{By Fold} & \multicolumn{2}{c}{Global} \\
\hline
ID & Variables & CV & RMSE & $R^2$& RMSE & $R^2$\\ \hline
1a &all & random  &  0.96 &0.87&0.97&0.86\\
1b& all & spatial  & \textbf{1.25}& \textbf{0.07} & \textbf{1.75}& \textbf{0.58}\\
2a& selected by RFE ``spatial`` & random  &0.93&0.88&0.95&0.87 \\
2b & selected by RFE ``spatial`` & spatial  & \textbf{1.22}& \textbf{0.06} & \textbf{1.73}& \textbf{0.58}\\
3a & selected by FFS random & random  & 0.91 &0.88 &0.92&0.88\\
3b & selected by FFS random & spatial &  \textbf{1.23} &  \textbf{0.04}& \textbf{1.80}& \textbf{0.58}\\
4a & selected by FFS spatial & random  &1.12&0.83&1.14&0.81\\
4b & selected by FFS spatial & spatial  &  \textbf{1.20} &  \textbf{0.06}& \textbf{1.64}& \textbf{0.63}\\
\hline
\end{tabular}
\end{table}

\subsection{Variable importance and selected variables}

\begin{figure}[htbp]
\includegraphics[width=1\textwidth]{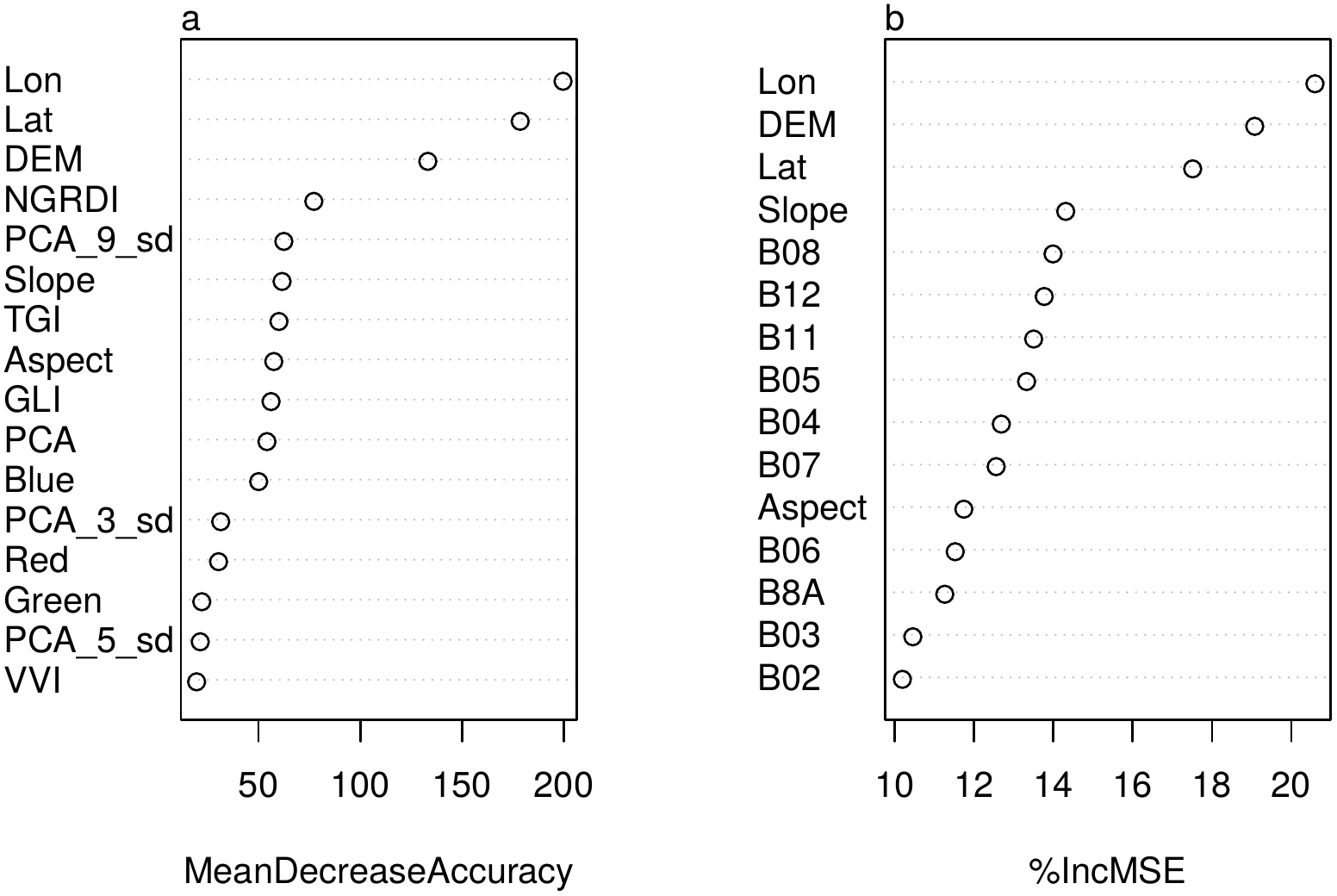}
\caption{Relative scaled importance of the predictor variables within the Random Forest models using all variables or using an RFE approach for the case study of predicting (a) LULC and (b) LAI. See section.~\ref{sec:LUC_predictors} and \ref{sec:LAI_predictors} for further explanations of the variables.}
\label{fig:varimp} 
\end{figure}

When all variables were presented to the algorithm to predict LULC and LAI, the most important variables were latitude, longitude and elevation (Fig.~\ref{fig:varimp}). The spectral predictor variables were considerably less important for these tasks.

The RFE based upon this variable importance ranking did not eliminate any variables for the LULC classification, hence the model is essentially identical to the initial full model. For the LAI prediction model, the RFE only dropped the Sentinel-2 band ``B02`` (blue band). The combination of FFS and random cross-validation selected latitude, longitude, DEM and aspect for LULC classification and DEM, longitude, latitude, B12, B07, aspect and slope for LAI predictions, in decreasing order of importance.

\begin{figure}[htb]
\includegraphics[width=1\textwidth]{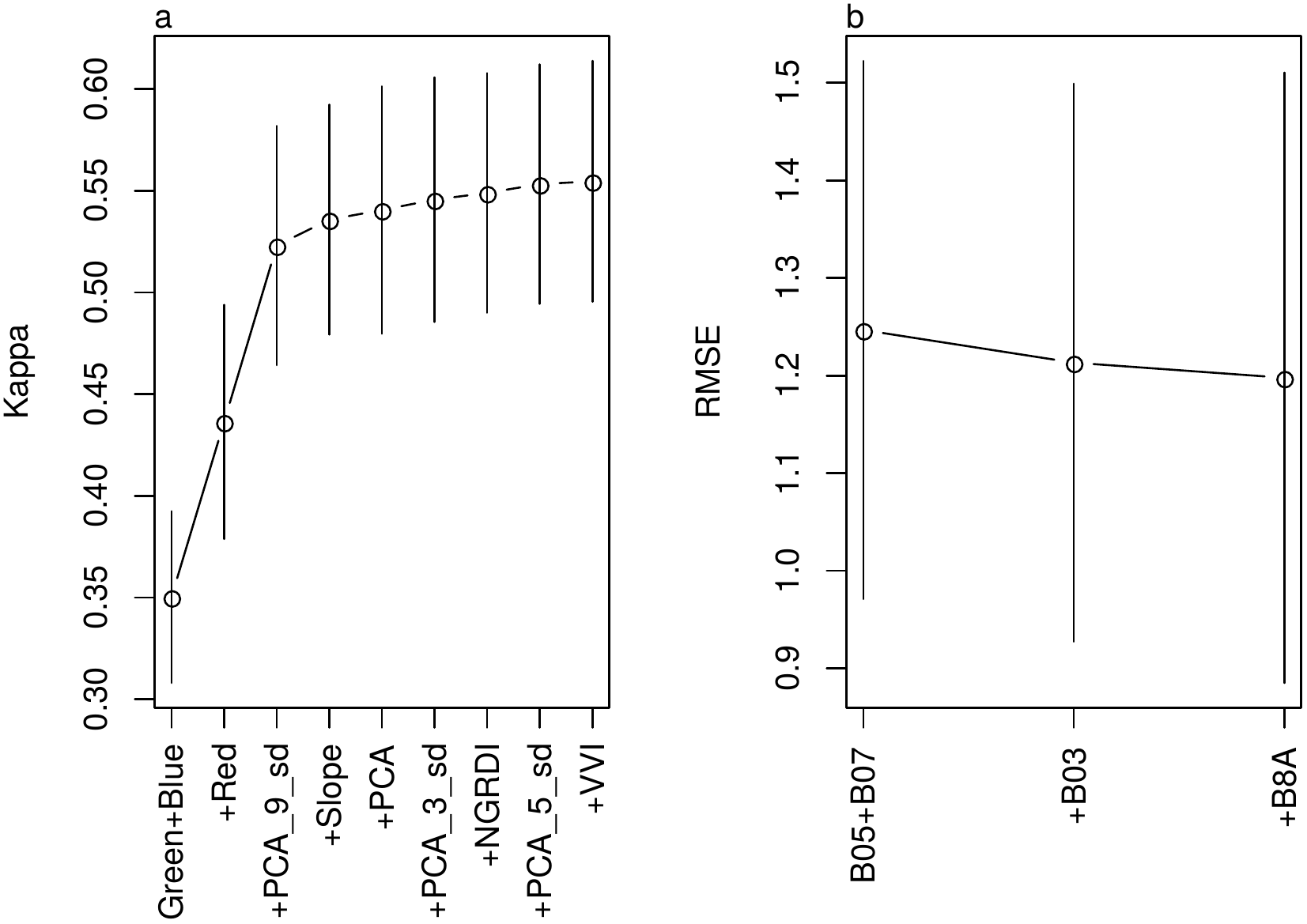}
\caption{Performance of FFS with variables selected by spatial cross-validation for prediction of (a) LULC and (b) LAI. The first point indicates the model’s performance using the two variables that lead to the best spatial performance. Subsequent points indicate the model’s performance with the addition of the next best variable, i.e. the third point represents the top four variables. Bars represent the standard deviation over the k spatial folds. See section.~\ref{sec:LUC_predictors} and \ref{sec:LAI_predictors} for further explanations of the variables.}
\label{fig:varimp_ffs} 
\end{figure}

FFS with spatial cross-validation identified the geographic coordinates and elevation as irrelevant or even counterproductive and dropped them from the models. The final model used only a subset of the spectral variables and the slope for the LULC classification. Here, green, blue, red and the standard deviation of the pca in a 9x9 environment made the largest contributions (Fig.~\ref{fig:varimp_ffs}a). For LAI predictions, only the bands B05, B07 (both red edge), B03 (green) and B8A (narrow NIR) were identified as important. However, B03 and B8A only slightly decreased the RMSE compared to the model that used B05 and B07 only (Fig.~\ref{fig:varimp_ffs}b).

\subsection{Visual assessment of the spatial prediction}

\begin{figure}[htbp]
\includegraphics[width=1\textwidth]{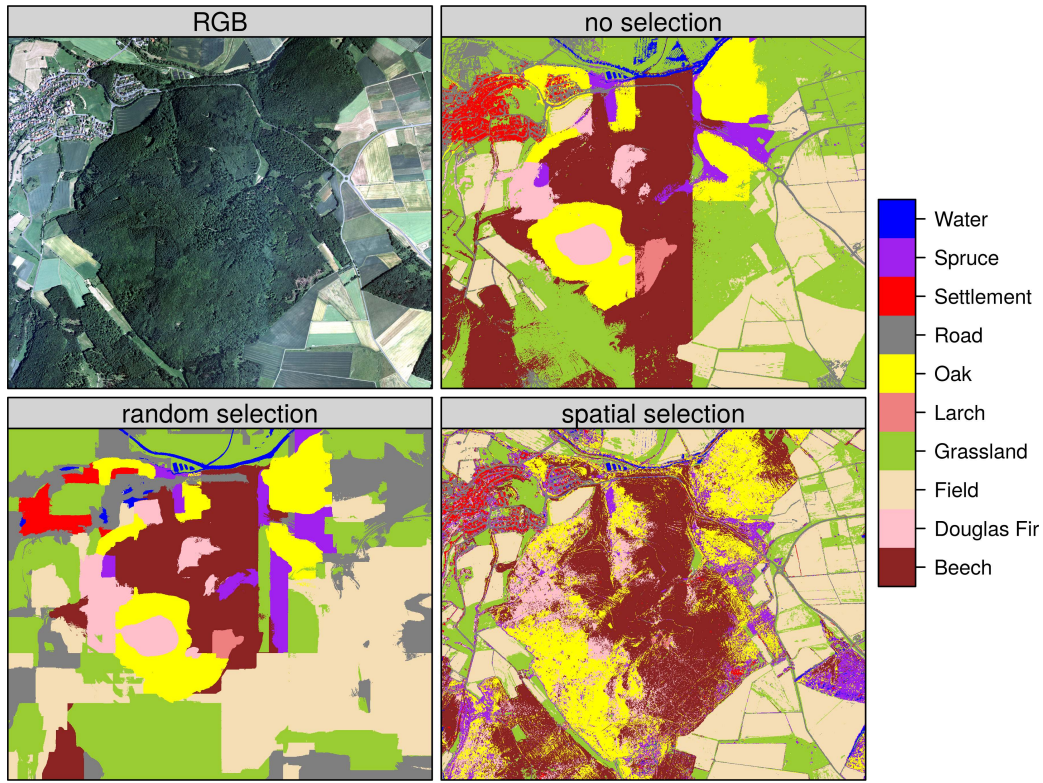}
\caption{RGB representation of the study area on the basis of the aerial image (RGB), spatial LULC predictions by the model that used all potential predictor variables (no selection), the model with variables being selected by random FFS (random selection), as well as spatial FFS (spatial selection).}
\label{fig:prediction_LUC} 
\end{figure}

The model that used all variables to predict LULC led to noticeably linear features when making spatial predictions for the full study area (Fig.~\ref{fig:prediction_LUC} no selection; the RFE-based model produces a quasi identical spatial prediction). A clear linear delineation was made between beech forest and grassland that does not correspond to the visual inspection of the underlying aerial image (Fig.~\ref{fig:prediction_LUC} RGB). An obvious patch of forest in the southeastern part of the image was absent from prediction, where it was classified as grassland. Field and road were clearly confused for one another in the northwestern corner. A round patch of Douglas fir in the southwestern quarter of the image can be clearly associated to the highest elevation of the forest. Elevation appeared to be an overwhelming factor in the prediction of water, since parts of fields in the north of the image corresponding to the lowest elevations (Fig.~\ref{fig:predictors}) were falsely classified as such. These areas were visually distinguishable from water in the RGB. Several other patterns that did not correspond to a visual interpretation of the RGB were also present.

Random variable selection (FFS with random selection) enhanced the problem and linear features became more obvious (Fig.~\ref{fig:prediction_LUC} random selection). The prediction has an overall smooth appearance as FFS removed spectral variables from the model.

When variables were selected by spatial FFS, the coordinates and elevation were removed by the algorithm and the classification was based on the spectral variables (Fig.~\ref{fig:prediction_LUC} spatial selection). The result showed much greater local variability that was clearly driven by the underlying spectral information rather than gradual changes driven by geolocation. No linear artefacts were observed. 

\begin{figure}[htbp]
\includegraphics[width=1\textwidth]{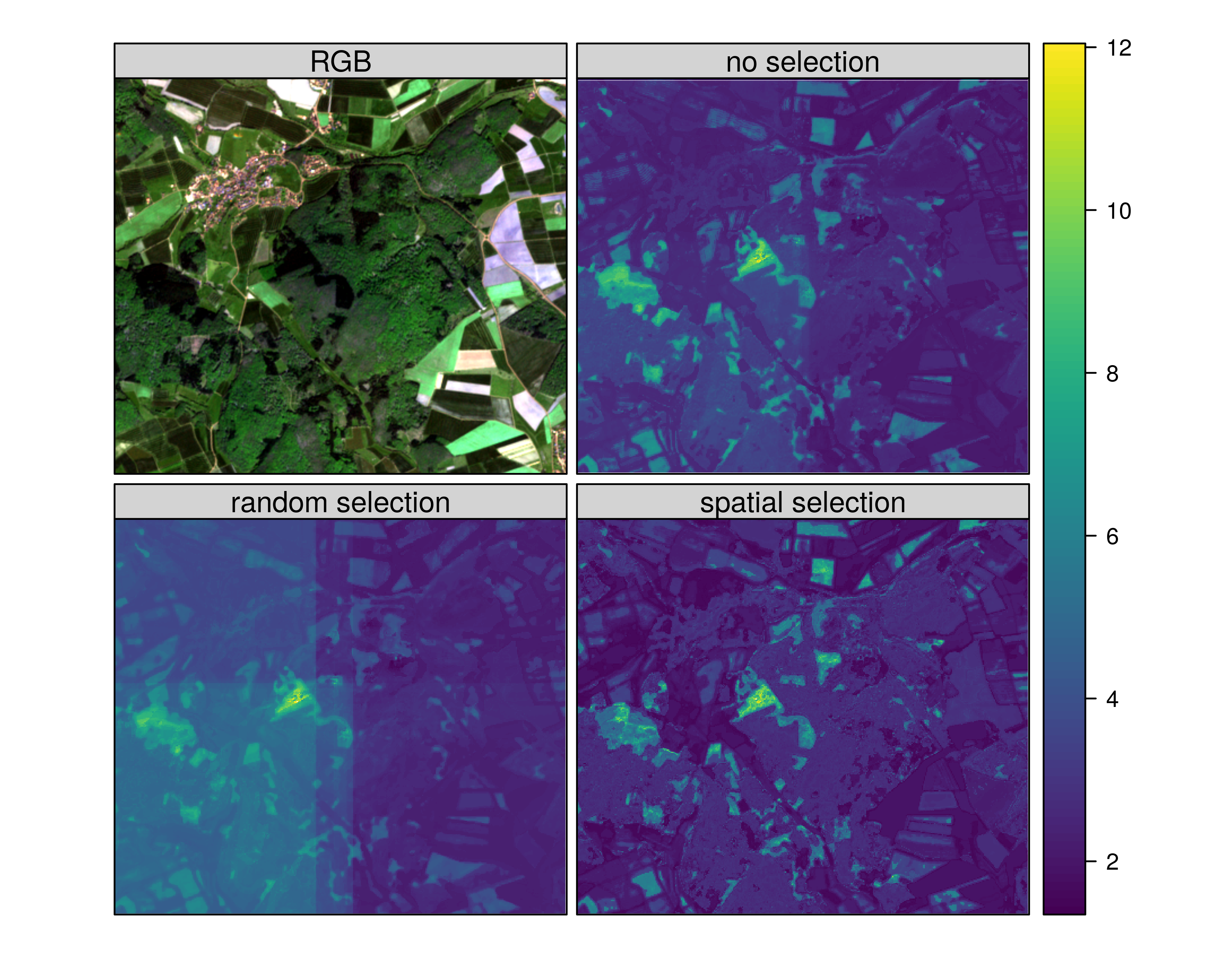}
\caption{RGB representation of the study area on the basis of the Sentinel-2 image (RGB), spatial LAI predictions by the model that used all potential predictor variables (no selection), the model with variables being selected by random FFS (random selection), as well as spatial FFS (spatial selection).}
\label{fig:prediction_LAI} 
\end{figure}

Similar though less striking patterns were found for the LAI predictions. Using all potential predictor variables led to a visible linear feature dividing generally lower LAI values to the east from generally higher values to the west (Fig.~\ref{fig:prediction_LAI} no selection). Such features became more obvious when FFS with random cross-validation was applied (Fig.~\ref{fig:prediction_LAI} random selection). When FFS selected variables with spatial cross-validation, no geolocation variables were selected and the results showed no obvious artefacts in the spatial prediction (Fig.~\ref{fig:prediction_LAI} spatial selection).

\section{Discussion}

\subsection{Importance of spatial validation}

The results clearly highlight again the necessity of spatial validation for realistically assessing the performance of spatial prediction models. Standard validation procedures that use random subsets of the dataset (i.e. random k-fold cross-validation) produce overoptimistic estimates about the model performance (in this case ``nearly perfect classification`` of LULC and low errors for LAI predictions). These do not provide information about the actual model performance with respect to the prediction of any other place than the sampling locations (i.e. create a map of LULC or LAI). Predicting any and not just the training locations, however, is the aim of most spatial prediction models, and the performance estimates must account for that which is increasingly and consistently recommended in recent literature \citep[e.g.][]{Wenger2012, Juel2015, Roberts2017, Valavi2018, Pohjankukka2017,Canovas-Garcia2017}. Random validation is not a meaningful strategy for spatial prediction tasks. This is especially essential when flexible algorithms are applied to highly clustered training samples that cause the risk of overfitting caused by the spatial autocorrelation of the predictor and response variables.

\subsection{Relevance of spatial variable selection}

It is not surprising that the model that uses all variable showed a strong performance in random cross-validation as compared to spatial cross-validation considering that within the polygons of digitized LULC or lidar-derived LAI training sites, the samples feature similar properties in their predictor variable space. Hence large parts of the training samples are not independent from one another. This leads to overfitting and incorrectly assigning high importance to the variables that represent the spatial location, which is especially clear for geolocation variables (i.e. latitude, longitude). Therefore, many studies have unsurprisingly identified coordinates as one of the, if not the, most important predictor, such as for tree species distribution \citep{Attorre2011}, monthly precipitation \citep{Jing2016}, deforestation \citep{Zanella2017}, phytoplankton abundance \citep{Roubeix2016} and explaining the spatial variability of soil organic carbon \citep{Yang2016}. According to our results, spatial variable selection would have very likely removed geolocation variables from these studies’ models. In addition to geolocation variables such as coordinates or Euclidean distance fields \citep{Behrens2018}, variables that are unique to a certain spatial cluster are also problematic. For example, \citep{Meyer2016c} show that elevation can complicate models when it is clearly indicative of one spatial cluster – in this study, circular patterns of Douglas fir were predicted for the areas of the forest with the greatest elevation, which is a clear artefact caused by misinterpretation of elevation as a predictor variable. Spatial FFS removed elevation as it was identified as unimportant or counterproductive as evinced by improved visual and statistical results.

We show that including location variables does not solve the problem of autocorrelation but intensifies the problem, at least for spatially clustered data. This finding contrasts with recommendations from previous studies \citep{Mascaro2013, Cracknell2014}. Though Random Forests are known for being robust to uninformative predictor variables, this study clearly shows that misleading variables can have negative effects on the models. This notion is also supported by \citet{Rocha2018}, who showed that spatial predictions of plant traits suffer when models include spatial relations. The phenomenon, however, can only be detected if spatial cross-validation is used. Spatial models evaluated in a default random way will still feign a high spatial performance.

Spatial cross-validation provides reliable performance measures for spatial models. However, it does not change the model itself. During the internal Random Forest training, variables are not selected by spatial cross-validation but by the out-of-bag (oob) error which is based on random bootstrapping. Hence in a Random Forest training, variables are not selected by their spatial importance. A spatial variable selection is therefore required to remove misleading variables from the models. Using geolocation variables, the algorithm could reproduce training samples with highest performance but only a selection of spatially meaningful variables allowed for predictions beyond the locations of the training samples. 

Having a look at the internal variable importance ranking of the Random Forests algorithm (Fig.~\ref{fig:varimp}) also explains why recursive feature selection cannot help, even if spatial cross-validation determines the optimal variables: Since RFE is based on the importance ranking of the variables, those that are misleading and responsible for overfitting are often highly ranked (e.g. latitude, longitude in this study) and hence not removed by RFE.

This study automatically analyzes which variables are misleading, counterproductive and cause overfitting and hence must be removed from the models. We show that removing these variables from the models improves the statistical spatial performance; a visual inspection also confirmed reliable patterns compared to the common model that uses all variables. The increase in statistical performance was less obvious than in \citet{Meyer2018}, who used spatio-temporal data that can be explained by a stronger autocorrelation due to the application to long time series. Therefore, improvements in performance will likely track with increasing degrees of autocorrelation if the sampling design includes clear spatial clusters.

The results of this study strongly suggest that spatial cross-validation needs to be considered not only for model validation and model tuning \citep[see][for a study on the relevance of spatial validation for hyperparameter tuning in machine learning applications] {Schratz2019} but also for variable selection, hence during all steps of model building.

\subsection{Need for visual assessment in addition to statistical validation}

The results also show that statistical validation alone is insufficient to validate spatial prediction models. Both the model using all potential predictors and the one using spatially selected variables perform statistically similar in spatial cross-validation. Hence, we conclude that the models perform equally well, statistically speaking. A visual assessment reveals that this assumption does not hold true, however. 
Removing misleading variables dramatically changes the actual outcome as patterns in the LULC and LAI predictions are considerably different.
Again, this highlights the need for spatial variable selection. In several other studies, artefacts are mainly visible as clear linear features and can most certainly be traced back to the geolocation variables. \citep{Jing2016} and \citep{Shi2015a} used coordinates for downscaling precipitation, which the Random Forest algorithm identified as the most important variable, but which resulted in visible linear patterns in the spatial prediction. Mud content prediction by \citet{Li2011} also shows linear patterns that are most likely caused by the inclusion of latitude and longitude as predictors. In a study by \citet{Fox2017} the Random Forest algorithm also ranked latitude and longitude as important, yet the resulting marine bird distribution along the Canadian coast shows clear linear cuts.
These examples visually highlight the issues that including geolocation variables can cause. They also underline the importance of spatial cross-validation for spatial error assessment in conjunction with spatial variable selection to ensure that only variables with actual predictive power beyond the training locations are included.

\section{Conclusions}

This study underlines the necessity of spatial validation strategies in spatial machine learning applications. Results will likely be overoptimistic if these strategies are ignored. This is especially the problem when there is strong spatial autocorrelation in the data and when training samples are clustered in space.

However, spatial machine learning applications should not be restricted to the usage of spatial validation. This study shows that certain variables are responsible for overfitting that causes strong random performance but a failure in predicting any other than the training samples location. This is especially evident for geolocation variables (e.g. latitude, longitude). When such variables are used in spatial modelling where training samples are highly clustered in space, they lead the algorithms to effectively reproduce the training data but lead the model to fail predicting on new samples. Hence, the Random Forest algorithm cannot interpret such variables in a meaningful way. Spatial variable selection is required to automatically select variables that are useful in a Random Forest setup, for example the suggested forward feature selection that selects variables according to their contribution for spatial predictions. Spatial validation should hence be considered during all steps of modelling, from hyperparameter tuning, variable selection to performance estimation.

Like most other machine learning algorithms, Random Forests have the reputation that no assumptions about the data distribution are necessary. However, the results of this study show that it might be necessary to revisit this idea and general guidelines should be formulated to make applications more objective.

Finally, the results of this study allow the conclusion that ignoring spatial dependencies in machine learning applications for spatial predictions carries a high risk of developing models that can reproduce training data well but do not make reliable spatial predictions. 
Reliable spatial predictions can only be achieved if spatial dependencies are taken into account during the modelling process, i.e. not only for the purpose of model validation, but also for the selection of appropriate predictor variables.


\section*{Acknowledgments}
This work was conducted within the Natur 4.0 | Sensing Biodiversity project funded by the Hessian state offensive for the development of scientific-economic excellence (LOEWE). 

\bibliography{Literature}

\begin{thebibliography}{53}
\expandafter\ifx\csname natexlab\endcsname\relax\def\natexlab#1{#1}\fi
\providecommand{\url}[1]{\texttt{#1}}
\providecommand{\href}[2]{#2}
\providecommand{\path}[1]{#1}
\providecommand{\DOIprefix}{doi:}
\providecommand{\ArXivprefix}{arXiv:}
\providecommand{\URLprefix}{URL: }
\providecommand{\Pubmedprefix}{pmid:}
\providecommand{\doi}[1]{\href{http://dx.doi.org/#1}{\path{#1}}}
\providecommand{\Pubmed}[1]{\href{pmid:#1}{\path{#1}}}
\providecommand{\bibinfo}[2]{#2}
\ifx\xfnm\relax \def\xfnm[#1]{\unskip,\space#1}\fi
\bibitem[{Appelhans et~al.(2015)Appelhans, Mwangomo, Hardy, Hemp \&
  Nauss}]{Appelhans2015}
\bibinfo{author}{Appelhans, T.}, \bibinfo{author}{Mwangomo, E.},
  \bibinfo{author}{Hardy, D.~R.}, \bibinfo{author}{Hemp, A.}, \&
  \bibinfo{author}{Nauss, T.} (\bibinfo{year}{2015}).
\newblock \bibinfo{title}{{E}valuating machine learning approaches for the
  interpolation of monthly air temperature at {M}t. {K}ilimanjaro, {T}anzania}.
\newblock {\it \bibinfo{journal}{Spatial Statistics}\/},  {\it
  \bibinfo{volume}{14, Part A}\/}, \bibinfo{pages}{91--113}. \URLprefix
  \url{https://doi.org/10.1016/j.spasta.2015.05.008}.
\bibitem[{Attorre et~al.(2011)Attorre, Alfò, Sanctis, Francesconi, Valenti,
  Vitale \& Bruno}]{Attorre2011}
\bibinfo{author}{Attorre, F.}, \bibinfo{author}{Alfò, M.},
  \bibinfo{author}{Sanctis, M.~D.}, \bibinfo{author}{Francesconi, F.},
  \bibinfo{author}{Valenti, R.}, \bibinfo{author}{Vitale, M.}, \&
  \bibinfo{author}{Bruno, F.} (\bibinfo{year}{2011}).
\newblock \bibinfo{title}{Evaluating the effects of climate change on tree
  species abundance and distribution in the italian peninsula}.
\newblock {\it \bibinfo{journal}{Appl. Veg. Sci.}\/},  {\it
  \bibinfo{volume}{14}\/}, \bibinfo{pages}{242--255}. \URLprefix
  \url{http://www.jstor.org/stable/41058163}.
\bibitem[{Bahn \& McGill(2013)}]{Bahn2013}
\bibinfo{author}{Bahn, V.}, \& \bibinfo{author}{McGill, B.~J.}
  (\bibinfo{year}{2013}).
\newblock \bibinfo{title}{Testing the predictive performance of distribution
  models}.
\newblock {\it \bibinfo{journal}{Oikos}\/},  {\it \bibinfo{volume}{122}\/},
  \bibinfo{pages}{321--331}. \URLprefix
  \url{https://onlinelibrary.wiley.com/doi/abs/10.1111/j.1600-0706.2012.00299.x}.
  \DOIprefix\doi{10.1111/j.1600-0706.2012.00299.x}.
  \href{http://arxiv.org/abs/https://onlinelibrary.wiley.com/doi/pdf/10.1111/j.1600-0706.2012.00299.x}{\tt
  arXiv:https://onlinelibrary.wiley.com/doi/pdf/10.1111/j.1600-0706.2012.00299.x}.
\bibitem[{Behrens et~al.(2018)Behrens, Schmidt, Viscarra~Rossel, Gries,
  Scholten \& MacMillan}]{Behrens2018}
\bibinfo{author}{Behrens, T.}, \bibinfo{author}{Schmidt, K.},
  \bibinfo{author}{Viscarra~Rossel, R.~A.}, \bibinfo{author}{Gries, P.},
  \bibinfo{author}{Scholten, T.}, \& \bibinfo{author}{MacMillan, R.~A.}
  (\bibinfo{year}{2018}).
\newblock \bibinfo{title}{Spatial modelling with euclidean distance fields and
  machine learning}.
\newblock {\it \bibinfo{journal}{Eur. J. Soil Sci.}\/},  {\it
  \bibinfo{volume}{69}\/}, \bibinfo{pages}{757--770}. \URLprefix
  \url{https://onlinelibrary.wiley.com/doi/abs/10.1111/ejss.12687}.
  \DOIprefix\doi{10.1111/ejss.12687}.
  \href{http://arxiv.org/abs/https://onlinelibrary.wiley.com/doi/pdf/10.1111/ejss.12687}{\tt
  arXiv:https://onlinelibrary.wiley.com/doi/pdf/10.1111/ejss.12687}.
\bibitem[{Breiman(2001)}]{Breiman2001}
\bibinfo{author}{Breiman, L.} (\bibinfo{year}{2001}).
\newblock \bibinfo{title}{{R}andom {F}orests}.
\newblock {\it \bibinfo{journal}{Machine Learning}\/},  {\it
  \bibinfo{volume}{45}\/}, \bibinfo{pages}{5--32}. \URLprefix
  \url{https://doi.org/10.1023/A:1010933404324}.
\bibitem[{Brenning(2005)}]{Brenning2005}
\bibinfo{author}{Brenning, A.} (\bibinfo{year}{2005}).
\newblock \bibinfo{title}{{S}patial prediction models for landslide hazards:
  review, comparison and evaluation}.
\newblock {\it \bibinfo{journal}{Nat. Hazards Earth Syst. Sci.}\/},  {\it
  \bibinfo{volume}{5}\/}, \bibinfo{pages}{853--862}. \URLprefix
  \url{https://doi.org/10.5194/nhess-5-853-2005}.
\bibitem[{Brungard et~al.(2015)Brungard, Boettinger, Duniway, Wills \&
  Jr.}]{Brungard2015}
\bibinfo{author}{Brungard, C.~W.}, \bibinfo{author}{Boettinger, J.~L.},
  \bibinfo{author}{Duniway, M.~C.}, \bibinfo{author}{Wills, S.~A.}, \&
  \bibinfo{author}{Jr., T. C.~E.} (\bibinfo{year}{2015}).
\newblock \bibinfo{title}{{M}achine learning for predicting soil classes in
  three semi-arid landscapes}.
\newblock {\it \bibinfo{journal}{Geoderma}\/},  {\it
  \bibinfo{volume}{239--240}\/}, \bibinfo{pages}{68--83}. \URLprefix
  \url{https://doi.org/10.1016/j.geoderma.2014.09.019}.
\bibitem[{C\'{a}novas-Garc\'{i}a et~al.(2017)C\'{a}novas-Garc\'{i}a,
  Alonso-Sarr\'{i}a, Gomariz-Castillo \& {n}ate
  Valdivieso}]{Canovas-Garcia2017}
\bibinfo{author}{C\'{a}novas-Garc\'{i}a, F.},
  \bibinfo{author}{Alonso-Sarr\'{i}a, F.}, \bibinfo{author}{Gomariz-Castillo,
  F.}, \& \bibinfo{author}{{n}ate Valdivieso, F.~O.} (\bibinfo{year}{2017}).
\newblock \bibinfo{title}{Modification of the random forest algorithm to avoid
  statistical dependence problems when classifying remote sensing imagery}.
\newblock {\it \bibinfo{journal}{Comput. Geosci.}\/},  {\it
  \bibinfo{volume}{103}\/}, \bibinfo{pages}{1 -- 11}. \URLprefix
  \url{http://www.sciencedirect.com/science/article/pii/S0098300416303909}.
  \DOIprefix\doi{https://doi.org/10.1016/j.cageo.2017.02.012}.
\bibitem[{Cohen(1960)}]{Cohen1960}
\bibinfo{author}{Cohen, J.} (\bibinfo{year}{1960}).
\newblock \bibinfo{title}{A coefficient of agreement for nominal scales}.
\newblock {\it \bibinfo{journal}{Educational and Psychological Measurement}\/},
   {\it \bibinfo{volume}{20}\/}, \bibinfo{pages}{37--46}. \URLprefix
  \url{https://doi.org/10.1177/001316446002000104}.
  \DOIprefix\doi{10.1177/001316446002000104}.
  \href{http://arxiv.org/abs/https://doi.org/10.1177/001316446002000104}{\tt
  arXiv:https://doi.org/10.1177/001316446002000104}.
\bibitem[{Cracknell \& Reading(2014)}]{Cracknell2014}
\bibinfo{author}{Cracknell, M.~J.}, \& \bibinfo{author}{Reading, A.~M.}
  (\bibinfo{year}{2014}).
\newblock \bibinfo{title}{{G}eological mapping using remote sensing data: {A}
  comparison of five machine learning algorithms, their response to variations
  in the spatial distribution of training data and the use of explicit spatial
  information}.
\newblock {\it \bibinfo{journal}{Computers \& Geosciences}\/},  {\it
  \bibinfo{volume}{63}\/}, \bibinfo{pages}{22--33}. \URLprefix
  \url{https://doi.org/10.1016/j.cageo.2013.10.008}.
\bibitem[{Fox et~al.(2017)Fox, Huettmann, Harvey, Morgan, Robinson, Williams \&
  Paquet}]{Fox2017}
\bibinfo{author}{Fox, C.~H.}, \bibinfo{author}{Huettmann, F.~H.},
  \bibinfo{author}{Harvey, G. K.~A.}, \bibinfo{author}{Morgan, K.~H.},
  \bibinfo{author}{Robinson, J.}, \bibinfo{author}{Williams, R.}, \&
  \bibinfo{author}{Paquet, P.~C.} (\bibinfo{year}{2017}).
\newblock \bibinfo{title}{Predictions from machine learning ensembles: marine
  bird distribution and density on canadas pacific coast}.
\newblock {\it \bibinfo{journal}{Mar. Ecol. Prog. Ser.}\/},  {\it
  \bibinfo{volume}{566}\/}, \bibinfo{pages}{199--216}. \URLprefix
  \url{https://www.int-res.com/abstracts/meps/v566/p199-216/}.
\bibitem[{Gasch et~al.(2015)Gasch, Hengl, Gr{\"a}ler, Meyer, Magney \&
  Brown}]{Gasch2015}
\bibinfo{author}{Gasch, C.~K.}, \bibinfo{author}{Hengl, T.},
  \bibinfo{author}{Gr{\"a}ler, B.}, \bibinfo{author}{Meyer, H.},
  \bibinfo{author}{Magney, T.~S.}, \& \bibinfo{author}{Brown, D.~J.}
  (\bibinfo{year}{2015}).
\newblock \bibinfo{title}{{S}patio-temporal interpolation of soil water,
  temperature, and electrical conductivity in 3{D} + {T}: {T}he {C}ook
  {A}gronomy {F}arm data set}.
\newblock {\it \bibinfo{journal}{Spatial Statistics}\/},  {\it
  \bibinfo{volume}{14, Part A}\/}, \bibinfo{pages}{70--90}.
\bibitem[{Georganos et~al.(2019)Georganos, Grippa, Gadiaga, Linard, Lennert,
  Vanhuysse, Mboga, Wolff \& Kalogirou}]{Georganos2019}
\bibinfo{author}{Georganos, S.}, \bibinfo{author}{Grippa, T.},
  \bibinfo{author}{Gadiaga, A.~N.}, \bibinfo{author}{Linard, C.},
  \bibinfo{author}{Lennert, M.}, \bibinfo{author}{Vanhuysse, S.},
  \bibinfo{author}{Mboga, N.}, \bibinfo{author}{Wolff, E.}, \&
  \bibinfo{author}{Kalogirou, S.} (\bibinfo{year}{2019}).
\newblock \bibinfo{title}{Geographical random forests: a spatial extension of
  the random forest algorithm to address spatial heterogeneity in remote
  sensing and population modelling}.
\newblock {\it \bibinfo{journal}{Geocarto International}\/},  {\it
  \bibinfo{volume}{0}\/}, \bibinfo{pages}{1--16}. \URLprefix
  \url{https://doi.org/10.1080/10106049.2019.1595177}.
  \DOIprefix\doi{10.1080/10106049.2019.1595177}.
  \href{http://arxiv.org/abs/https://doi.org/10.1080/10106049.2019.1595177}{\tt
  arXiv:https://doi.org/10.1080/10106049.2019.1595177}.
\bibitem[{Getzin et~al.(2017)Getzin, Fischer, Knapp \& Huth}]{Getzin2017}
\bibinfo{author}{Getzin, S.}, \bibinfo{author}{Fischer, R.},
  \bibinfo{author}{Knapp, N.}, \& \bibinfo{author}{Huth, A.}
  (\bibinfo{year}{2017}).
\newblock \bibinfo{title}{Using airborne lidar to assess spatial heterogeneity
  in forest structure on mount kilimanjaro}.
\newblock {\it \bibinfo{journal}{Landscape Ecol.}\/},  {\it
  \bibinfo{volume}{32}\/}, \bibinfo{pages}{1881--1894}. \URLprefix
  \url{https://doi.org/10.1007/s10980-017-0550-7}.
  \DOIprefix\doi{10.1007/s10980-017-0550-7}.
\bibitem[{Ghosh \& Joshi(2014)}]{Ghosh2014}
\bibinfo{author}{Ghosh, A.}, \& \bibinfo{author}{Joshi, P.}
  (\bibinfo{year}{2014}).
\newblock \bibinfo{title}{{A} comparison of selected classification algorithms
  for mapping bamboo patches in lower {G}angetic plains using very high
  resolution {W}orld{V}iew 2 imagery}.
\newblock {\it \bibinfo{journal}{Int. J. Appl. Earth Obs. Geoinf.}\/},  {\it
  \bibinfo{volume}{26}\/}, \bibinfo{pages}{298--311}. \URLprefix
  \url{https://doi.org/10.1016/j.jag.2013.08.011}.
\bibitem[{Gudmundsson \& Seneviratne(2015)}]{Gudmundsson2015}
\bibinfo{author}{Gudmundsson, L.}, \& \bibinfo{author}{Seneviratne, S.~I.}
  (\bibinfo{year}{2015}).
\newblock \bibinfo{title}{{T}owards observation-based gridded runoff estimates
  for {E}urope}.
\newblock {\it \bibinfo{journal}{Hydrol. Earth Syst. Sci.}\/},  {\it
  \bibinfo{volume}{19}\/}, \bibinfo{pages}{2859--2879}.
\bibitem[{Hengl et~al.(2018)Hengl, Nussbaum, Wright, Heuvelink \&
  Gräler}]{Hengl2018}
\bibinfo{author}{Hengl, T.}, \bibinfo{author}{Nussbaum, M.},
  \bibinfo{author}{Wright, M.}, \bibinfo{author}{Heuvelink, G.}, \&
  \bibinfo{author}{Gräler, B.} (\bibinfo{year}{2018}).
\newblock \bibinfo{title}{Random forest as a generic framework for predictive
  modeling of spatial and spatio-temporal variables}.
\newblock {\it \bibinfo{journal}{PeerJ Preprints}\/}, .
  \DOIprefix\doi{10.7287/peerj.preprints.26693v3}.
\bibitem[{{Hessische Verwaltung f{\"u}r Bodenmanagement und
  Geoinformation}(2018{\natexlab{a}})}]{HVBG2018a}
\bibinfo{author}{{Hessische Verwaltung f{\"u}r Bodenmanagement und
  Geoinformation}} (\bibinfo{year}{2018}{\natexlab{a}}).
\newblock \bibinfo{title}{Aerial imagery}.
\bibitem[{{Hessische Verwaltung f{\"u}r Bodenmanagement und
  Geoinformation}(2018{\natexlab{b}})}]{HVBG2018}
\bibinfo{author}{{Hessische Verwaltung f{\"u}r Bodenmanagement und
  Geoinformation}} (\bibinfo{year}{2018}{\natexlab{b}}).
\newblock \bibinfo{title}{Lidar data}.
\bibitem[{Hunt et~al.(2013)Hunt, Doraiswamy, McMurtrey, Daughtry, Perry \&
  Akhmedov}]{Hunt2013}
\bibinfo{author}{Hunt, E.~R.}, \bibinfo{author}{Doraiswamy, P.~C.},
  \bibinfo{author}{McMurtrey, J.~E.}, \bibinfo{author}{Daughtry, C.~S.},
  \bibinfo{author}{Perry, E.~M.}, \& \bibinfo{author}{Akhmedov, B.}
  (\bibinfo{year}{2013}).
\newblock \bibinfo{title}{A visible band index for remote sensing leaf
  chlorophyll content at the canopy scale}.
\newblock {\it \bibinfo{journal}{Int. J. Appl. Earth Obs. Geoinf.}\/},  {\it
  \bibinfo{volume}{21}\/}, \bibinfo{pages}{103 -- 112}. \URLprefix
  \url{http://www.sciencedirect.com/science/article/pii/S0303243412001791}.
  \DOIprefix\doi{https://doi.org/10.1016/j.jag.2012.07.020}.
\bibitem[{Janatian et~al.(2017)Janatian, Sadeghi, Sanaeinejad, Bakhshian,
  Farid, Hasheminia \& Ghazanfari}]{Janatian2017}
\bibinfo{author}{Janatian, N.}, \bibinfo{author}{Sadeghi, M.},
  \bibinfo{author}{Sanaeinejad, S.~H.}, \bibinfo{author}{Bakhshian, E.},
  \bibinfo{author}{Farid, A.}, \bibinfo{author}{Hasheminia, S.~M.}, \&
  \bibinfo{author}{Ghazanfari, S.} (\bibinfo{year}{2017}).
\newblock \bibinfo{title}{{A} statistical framework for estimating air
  temperature using {MODIS} land surface temperature data}.
\newblock {\it \bibinfo{journal}{Int. J. Climatol.}\/},  {\it
  \bibinfo{volume}{37}\/}, \bibinfo{pages}{1181--1194}. \URLprefix
  \url{https://doi.org/10.1002/joc.4766}.
\bibitem[{Jing et~al.(2016)Jing, Yang, Yue \& Zhao}]{Jing2016}
\bibinfo{author}{Jing, W.}, \bibinfo{author}{Yang, Y.}, \bibinfo{author}{Yue,
  X.}, \& \bibinfo{author}{Zhao, X.} (\bibinfo{year}{2016}).
\newblock \bibinfo{title}{{A} {C}omparison of {D}ifferent {R}egression
  {A}lgorithms for {D}ownscaling {M}onthly {S}atellite-{B}ased {P}recipitation
  over {N}orth {C}hina}.
\newblock {\it \bibinfo{journal}{Remote Sensing}\/},  {\it
  \bibinfo{volume}{8}\/}, \bibinfo{pages}{835}. \URLprefix
  \url{https://doi.org/10.3390/rs8100835}.
\bibitem[{Juel et~al.(2015)Juel, Groom, Svenning \& Ejrnæs}]{Juel2015}
\bibinfo{author}{Juel, A.}, \bibinfo{author}{Groom, G.~B.},
  \bibinfo{author}{Svenning, J.-C.}, \& \bibinfo{author}{Ejrnæs, R.}
  (\bibinfo{year}{2015}).
\newblock \bibinfo{title}{{S}patial application of {R}andom {F}orest models for
  fine-scale coastal vegetation classification using object based analysis of
  aerial orthophoto and {DEM} data}.
\newblock {\it \bibinfo{journal}{Int. J. Appl. Earth Obs. Geoinf.}\/},  {\it
  \bibinfo{volume}{42}\/}, \bibinfo{pages}{106 -- 114}. \URLprefix
  \url{http://www.sciencedirect.com/science/article/pii/S0303243415001178}.
  \DOIprefix\doi{https://doi.org/10.1016/j.jag.2015.05.008}.
\bibitem[{Kuhn(2016)}]{Kuhn2016}
\bibinfo{author}{Kuhn, M.} (\bibinfo{year}{2016}).
\newblock {\it \bibinfo{title}{caret: {C}lassification and {R}egression
  {T}raining}\/}.
\newblock \URLprefix \url{https://CRAN.R-project.org/package=caret}
  \bibinfo{note}{{R} package version 6.0-68}.
\bibitem[{Kuhn \& Johnson(2013)}]{Kuhn2013}
\bibinfo{author}{Kuhn, M.}, \& \bibinfo{author}{Johnson, K.}
  (\bibinfo{year}{2013}).
\newblock {\it \bibinfo{title}{{A}pplied {P}redictive {M}odeling}\/}.
\newblock (\bibinfo{edition}{1st} ed.).
\newblock \bibinfo{address}{New York}: \bibinfo{publisher}{Springer}.
\bibitem[{Langella et~al.(2010)Langella, Basile, Bonfante \&
  Terribile}]{Langella2010}
\bibinfo{author}{Langella, G.}, \bibinfo{author}{Basile, A.},
  \bibinfo{author}{Bonfante, A.}, \& \bibinfo{author}{Terribile, F.}
  (\bibinfo{year}{2010}).
\newblock \bibinfo{title}{{H}igh-resolution space-time rainfall analysis using
  integrated {ANN} inference systems}.
\newblock {\it \bibinfo{journal}{J. Hydrol.}\/},  {\it
  \bibinfo{volume}{387}\/}, \bibinfo{pages}{328--342}.
\bibitem[{Lary et~al.(2016)Lary, Alavi, Gandomi \& Walker}]{Lary2016}
\bibinfo{author}{Lary, D.~J.}, \bibinfo{author}{Alavi, A.~H.},
  \bibinfo{author}{Gandomi, A.~H.}, \& \bibinfo{author}{Walker, A.~L.}
  (\bibinfo{year}{2016}).
\newblock \bibinfo{title}{{M}achine learning in geosciences and remote
  sensing}.
\newblock {\it \bibinfo{journal}{Geosci. Front.}\/},  {\it
  \bibinfo{volume}{7}\/}, \bibinfo{pages}{3--10}. \URLprefix
  \url{https://doi.org/10.1016/j.gsf.2015.07.003}.
\bibitem[{Le~Rest et~al.(2014)Le~Rest, Pinaud, Monestiez, Chadoeuf \&
  Bretagnolle}]{LeRest2014}
\bibinfo{author}{Le~Rest, K.}, \bibinfo{author}{Pinaud, D.},
  \bibinfo{author}{Monestiez, P.}, \bibinfo{author}{Chadoeuf, J.}, \&
  \bibinfo{author}{Bretagnolle, V.} (\bibinfo{year}{2014}).
\newblock \bibinfo{title}{{S}patial leave-one-out cross-validation for variable
  selection in the presence of spatial autocorrelation}.
\newblock {\it \bibinfo{journal}{Global Ecol. Biogeogr.}\/},  {\it
  \bibinfo{volume}{23}\/}, \bibinfo{pages}{811--820}. \URLprefix
  \url{http://dx.doi.org/10.1111/geb.12161}. \DOIprefix\doi{10.1111/geb.12161}.
\bibitem[{Li et~al.(2011)Li, Heap, Potter \& Daniell}]{Li2011}
\bibinfo{author}{Li, J.}, \bibinfo{author}{Heap, A.~D.},
  \bibinfo{author}{Potter, A.}, \& \bibinfo{author}{Daniell, J.~J.}
  (\bibinfo{year}{2011}).
\newblock \bibinfo{title}{{A}pplication of machine learning methods to spatial
  interpolation of environmental variables}.
\newblock {\it \bibinfo{journal}{Environmental Modelling \& Software}\/},  {\it
  \bibinfo{volume}{26}\/}, \bibinfo{pages}{1647--1659}.
\bibitem[{Liaw \& Wiener(2002)}]{Liaw2002}
\bibinfo{author}{Liaw, A.}, \& \bibinfo{author}{Wiener, M.}
  (\bibinfo{year}{2002}).
\newblock \bibinfo{title}{{C}lassification and {R}egression by random{F}orest}.
\newblock {\it \bibinfo{journal}{R News}\/},  {\it \bibinfo{volume}{2}\/},
  \bibinfo{pages}{18--22}.
\bibitem[{Mascaro et~al.(2013)Mascaro, Asner, Knapp, Kennedy-Bowdoin, Martin,
  Anderson, Higgins \& Chadwick}]{Mascaro2013}
\bibinfo{author}{Mascaro, J.}, \bibinfo{author}{Asner, G.~P.},
  \bibinfo{author}{Knapp, D.~E.}, \bibinfo{author}{Kennedy-Bowdoin, T.},
  \bibinfo{author}{Martin, R.~E.}, \bibinfo{author}{Anderson, C.},
  \bibinfo{author}{Higgins, M.}, \& \bibinfo{author}{Chadwick, K.~D.}
  (\bibinfo{year}{2013}).
\newblock \bibinfo{title}{{A} {T}ale of {T}wo “{F}orests”: {R}andom
  {F}orest {M}achine {L}earning {A}ids {T}ropical {F}orest {C}arbon {M}apping}.
\newblock {\it \bibinfo{journal}{PLoS One}\/},  {\it \bibinfo{volume}{9}\/},
  \bibinfo{pages}{e85993--}. \URLprefix
  \url{http://www.ncbi.nlm.nih.gov/pmc/articles/PMC3904849/}.
\bibitem[{Meyer(2018)}]{Meyer2018c}
\bibinfo{author}{Meyer, H.} (\bibinfo{year}{2018}).
\newblock {\it \bibinfo{title}{CAST: 'caret' Applications for Spatial-Temporal
  Models}\/}.
\newblock \URLprefix \url{https://CRAN.R-project.org/package=CAST}
  \bibinfo{note}{r package version 0.2.1}.
\bibitem[{Meyer et~al.(2016)Meyer, Katurji, Appelhans, M{\"u}ller, Nauss,
  Roudier \& Zawar-Reza}]{Meyer2016c}
\bibinfo{author}{Meyer, H.}, \bibinfo{author}{Katurji, M.},
  \bibinfo{author}{Appelhans, T.}, \bibinfo{author}{M{\"u}ller, M.~U.},
  \bibinfo{author}{Nauss, T.}, \bibinfo{author}{Roudier, P.}, \&
  \bibinfo{author}{Zawar-Reza, P.} (\bibinfo{year}{2016}).
\newblock \bibinfo{title}{{M}apping {D}aily {A}ir {T}emperature for
  {A}ntarctica {B}ased on {MODIS} {LST}}.
\newblock {\it \bibinfo{journal}{Remote Sensing}\/},  {\it
  \bibinfo{volume}{8}\/}, \bibinfo{pages}{732}. \URLprefix
  \url{https://doi.org/10.3390/rs8090732}.
\bibitem[{Meyer et~al.(2017{\natexlab{a}})Meyer, K{\"u}hnlein, Reudenbach \&
  Nauss}]{Meyer2017a}
\bibinfo{author}{Meyer, H.}, \bibinfo{author}{K{\"u}hnlein, M.},
  \bibinfo{author}{Reudenbach, C.}, \& \bibinfo{author}{Nauss, T.}
  (\bibinfo{year}{2017}{\natexlab{a}}).
\newblock \bibinfo{title}{{R}evealing the potential of spectral and textural
  predictor variables in a neural network-based rainfall retrieval technique}.
\newblock {\it \bibinfo{journal}{Remote Sensing Letters}\/},  {\it
  \bibinfo{volume}{8}\/}, \bibinfo{pages}{647--656}. \URLprefix
  \url{https://doi.org/10.1080/2150704X.2017.1312026}.
\bibitem[{Meyer et~al.(2017{\natexlab{b}})Meyer, Lehnert, Wang, Reudenbach,
  Nauss \& Bendix}]{Meyer2017}
\bibinfo{author}{Meyer, H.}, \bibinfo{author}{Lehnert, L.~W.},
  \bibinfo{author}{Wang, Y.}, \bibinfo{author}{Reudenbach, C.},
  \bibinfo{author}{Nauss, T.}, \& \bibinfo{author}{Bendix, J.}
  (\bibinfo{year}{2017}{\natexlab{b}}).
\newblock \bibinfo{title}{{F}rom local spectral measurements to maps of
  vegetation cover and biomass on the {Q}inghai-{T}ibet-{P}lateau: {D}o we need
  hyperspectral information?}
\newblock {\it \bibinfo{journal}{Int. J. Appl. Earth Obs. Geoinf.}\/},  {\it
  \bibinfo{volume}{55}\/}, \bibinfo{pages}{21--31}. \URLprefix
  \url{https://doi.org/10.1016/j.jag.2016.10.001}.
\bibitem[{Meyer et~al.(2018)Meyer, Reudenbach, Hengl, Katurji \&
  Nauss}]{Meyer2018}
\bibinfo{author}{Meyer, H.}, \bibinfo{author}{Reudenbach, C.},
  \bibinfo{author}{Hengl, T.}, \bibinfo{author}{Katurji, M.}, \&
  \bibinfo{author}{Nauss, T.} (\bibinfo{year}{2018}).
\newblock \bibinfo{title}{{I}mproving performance of spatio-temporal machine
  learning models using forward feature selection and target-oriented
  validation}.
\newblock {\it \bibinfo{journal}{Environmental Modelling \& Software}\/},  {\it
  \bibinfo{volume}{101}\/}, \bibinfo{pages}{1 -- 9}. \URLprefix
  \url{https://doi.org/10.1016/j.envsoft.2017.12.001}.
  \DOIprefix\doi{https://doi.org/10.1016/j.envsoft.2017.12.001}.
\bibitem[{Micheletti et~al.(2014)Micheletti, Foresti, Robert, Leuenberger,
  Pedrazzini, Jaboyedoff \& Kanevski}]{Micheletti2014}
\bibinfo{author}{Micheletti, N.}, \bibinfo{author}{Foresti, L.},
  \bibinfo{author}{Robert, S.}, \bibinfo{author}{Leuenberger, M.},
  \bibinfo{author}{Pedrazzini, A.}, \bibinfo{author}{Jaboyedoff, M.}, \&
  \bibinfo{author}{Kanevski, M.} (\bibinfo{year}{2014}).
\newblock \bibinfo{title}{{M}achine {L}earning {F}eature {S}election {M}ethods
  for {L}andslide {S}usceptibility {M}apping}.
\newblock {\it \bibinfo{journal}{Math. Geosci.}\/},  {\it
  \bibinfo{volume}{46}\/}, \bibinfo{pages}{33--57}.
\bibitem[{{Planetary Habitability Laboratory}(2015)}]{PHL2015}
\bibinfo{author}{{Planetary Habitability Laboratory}} (\bibinfo{year}{2015}).
\newblock \bibinfo{title}{Visible vegetation index (vvi)}.
\newblock \URLprefix
  \url{http://phl.upr.edu/projects/visible-vegetation-index-vvi}.
\bibitem[{Pohjankukka et~al.(2017)Pohjankukka, Pahikkala, Nevalainen \&
  Heikkonen}]{Pohjankukka2017}
\bibinfo{author}{Pohjankukka, J.}, \bibinfo{author}{Pahikkala, T.},
  \bibinfo{author}{Nevalainen, P.}, \& \bibinfo{author}{Heikkonen, J.}
  (\bibinfo{year}{2017}).
\newblock \bibinfo{title}{{E}stimating the prediction performance of spatial
  models via spatial k-fold cross validation}.
\newblock {\it \bibinfo{journal}{International Journal of Geographical
  Information Science}\/},  {\it \bibinfo{volume}{31}\/},
  \bibinfo{pages}{2001--2019}. \URLprefix
  \url{https://doi.org/10.1080/13658816.2017.1346255}.
  \DOIprefix\doi{10.1080/13658816.2017.1346255}.
  \href{http://arxiv.org/abs/https://doi.org/10.1080/13658816.2017.1346255}{\tt
  arXiv:https://doi.org/10.1080/13658816.2017.1346255}.
\bibitem[{{R Core Team}(2018)}]{RCT2018}
\bibinfo{author}{{R Core Team}} (\bibinfo{year}{2018}).
\newblock {\it \bibinfo{title}{R: A Language and Environment for Statistical
  Computing}\/}.
\newblock \bibinfo{organization}{R Foundation for Statistical Computing}
  \bibinfo{address}{Vienna, Austria}.
\newblock \URLprefix \url{https://www.R-project.org/}.
\bibitem[{Roberts et~al.(2017)Roberts, Bahn, Ciuti, Boyce, Elith,
  Guillera-Arroita, Hauenstein, Lahoz-Monfort, Schr{\"o}der, Thuiller, Warton,
  Wintle, Hartig \& Dormann}]{Roberts2017}
\bibinfo{author}{Roberts, D.~R.}, \bibinfo{author}{Bahn, V.},
  \bibinfo{author}{Ciuti, S.}, \bibinfo{author}{Boyce, M.~S.},
  \bibinfo{author}{Elith, J.}, \bibinfo{author}{Guillera-Arroita, G.},
  \bibinfo{author}{Hauenstein, S.}, \bibinfo{author}{Lahoz-Monfort, J.~J.},
  \bibinfo{author}{Schr{\"o}der, B.}, \bibinfo{author}{Thuiller, W.},
  \bibinfo{author}{Warton, D.~I.}, \bibinfo{author}{Wintle, B.~A.},
  \bibinfo{author}{Hartig, F.}, \& \bibinfo{author}{Dormann, C.~F.}
  (\bibinfo{year}{2017}).
\newblock \bibinfo{title}{{C}ross-validation strategies for data with temporal,
  spatial, hierarchical, or phylogenetic structure}.
\newblock {\it \bibinfo{journal}{Ecography}\/}, .
  \DOIprefix\doi{10.1111/ecog.02881}.
\bibitem[{Rocha et~al.(2018)Rocha, Groen, Skidmore, Darvishzadeh \&
  Willemen}]{Rocha2018}
\bibinfo{author}{Rocha, A.~D.}, \bibinfo{author}{Groen, T.~A.},
  \bibinfo{author}{Skidmore, A.~K.}, \bibinfo{author}{Darvishzadeh, R.}, \&
  \bibinfo{author}{Willemen, L.} (\bibinfo{year}{2018}).
\newblock \bibinfo{title}{Machine learning using hyperspectral data
  inaccurately predicts plant traits under spatial dependency}.
\newblock {\it \bibinfo{journal}{Remote Sensing}\/},  {\it
  \bibinfo{volume}{10}\/}. \URLprefix
  \url{http://www.mdpi.com/2072-4292/10/8/1263}.
  \DOIprefix\doi{10.3390/rs10081263}.
\bibitem[{Roubeix et~al.(2016)Roubeix, Danis, Feret \& Baudoin}]{Roubeix2016}
\bibinfo{author}{Roubeix, V.}, \bibinfo{author}{Danis, P.-A.},
  \bibinfo{author}{Feret, T.}, \& \bibinfo{author}{Baudoin, J.-M.}
  (\bibinfo{year}{2016}).
\newblock \bibinfo{title}{Identification of ecological thresholds from
  variations in phytoplankton communities among lakes: contribution to the
  definition of environmental standards}.
\newblock {\it \bibinfo{journal}{Environ. Monit. Assess.}\/},  {\it
  \bibinfo{volume}{188}\/}, \bibinfo{pages}{246}. \URLprefix
  \url{https://doi.org/10.1007/s10661-016-5238-y}.
  \DOIprefix\doi{10.1007/s10661-016-5238-y}.
\bibitem[{Schratz et~al.(2019)Schratz, Muenchow, Iturritxa, Richter \&
  Brenning}]{Schratz2019}
\bibinfo{author}{Schratz, P.}, \bibinfo{author}{Muenchow, J.},
  \bibinfo{author}{Iturritxa, E.}, \bibinfo{author}{Richter, J.}, \&
  \bibinfo{author}{Brenning, A.} (\bibinfo{year}{2019}).
\newblock \bibinfo{title}{Hyperparameter tuning and performance assessment of
  statistical and machine-learning algorithms using spatial data}.
\newblock {\it \bibinfo{journal}{Ecological Modelling}\/},  {\it
  \bibinfo{volume}{406}\/}, \bibinfo{pages}{109 -- 120}. \URLprefix
  \url{http://www.sciencedirect.com/science/article/pii/S0304380019302145}.
  \DOIprefix\doi{https://doi.org/10.1016/j.ecolmodel.2019.06.002}.
\bibitem[{Shi et~al.(2015)Shi, Song, Xia, Lin, Myneni, Choi, Wang, Ni, Lao \&
  Yang}]{Shi2015a}
\bibinfo{author}{Shi, Y.}, \bibinfo{author}{Song, L.}, \bibinfo{author}{Xia,
  Z.}, \bibinfo{author}{Lin, Y.}, \bibinfo{author}{Myneni, R.~B.},
  \bibinfo{author}{Choi, S.}, \bibinfo{author}{Wang, L.}, \bibinfo{author}{Ni,
  X.}, \bibinfo{author}{Lao, C.}, \& \bibinfo{author}{Yang, F.}
  (\bibinfo{year}{2015}).
\newblock \bibinfo{title}{{M}apping {A}nnual {P}recipitation across {M}ainland
  {C}hina in the {P}eriod 2001-2010 from {TRMM}3{B}43 {P}roduct {U}sing
  {S}patial {D}ownscaling {A}pproach}.
\newblock {\it \bibinfo{journal}{Remote Sensing}\/},  {\it
  \bibinfo{volume}{7}\/}, \bibinfo{pages}{5849--5878}. \URLprefix
  \url{https://doi.org/10.3390/rs70505849}.
\bibitem[{Stevens et~al.(2013)Stevens, Nocita, T{\'o}th, Montanarella \& van
  Wesemael}]{Stevens2013}
\bibinfo{author}{Stevens, A.}, \bibinfo{author}{Nocita, M.},
  \bibinfo{author}{T{\'o}th, G.}, \bibinfo{author}{Montanarella, L.}, \&
  \bibinfo{author}{van Wesemael, B.} (\bibinfo{year}{2013}).
\newblock \bibinfo{title}{{P}rediction of {S}oil {O}rganic {C}arbon at the
  {E}uropean {S}cale by {V}isible and {N}ear {I}nfra{R}ed {R}eflectance
  {S}pectroscopy}.
\newblock {\it \bibinfo{journal}{PLoS One}\/},  {\it \bibinfo{volume}{8}\/},
  \bibinfo{pages}{1--13}. \URLprefix
  \url{https://doi.org/10.1371/journal.pone.0066409}.
\bibitem[{Valavi et~al.(2018)Valavi, Elith, Lahoz-Monfort \&
  Guillera-Arroita}]{Valavi2018}
\bibinfo{author}{Valavi, R.}, \bibinfo{author}{Elith, J.},
  \bibinfo{author}{Lahoz-Monfort, J.~J.}, \& \bibinfo{author}{Guillera-Arroita,
  G.} (\bibinfo{year}{2018}).
\newblock \bibinfo{title}{blockcv: an r package for generating spatially or
  environmentally separated folds for k-fold cross-validation of species
  distribution models}.
\newblock {\it \bibinfo{journal}{bioRxiv}\/}, . \URLprefix
  \url{https://www.biorxiv.org/content/early/2018/06/28/357798}.
  \DOIprefix\doi{10.1101/357798}.
  \href{http://arxiv.org/abs/https://www.biorxiv.org/content/early/2018/06/28/357798.full.pdf}{\tt
  arXiv:https://www.biorxiv.org/content/early/2018/06/28/357798.full.pdf}.
\bibitem[{Walsh et~al.(2017)Walsh, Kreakie, Cantwell \& Nacci}]{Walsh2017}
\bibinfo{author}{Walsh, E.~S.}, \bibinfo{author}{Kreakie, B.~J.},
  \bibinfo{author}{Cantwell, M.~G.}, \& \bibinfo{author}{Nacci, D.}
  (\bibinfo{year}{2017}).
\newblock \bibinfo{title}{A random forest approach to predict the spatial
  distribution of sediment pollution in an estuarine system}.
\newblock {\it \bibinfo{journal}{PLoS One}\/},  {\it \bibinfo{volume}{12}\/},
  \bibinfo{pages}{1--18}. \URLprefix
  \url{https://doi.org/10.1371/journal.pone.0179473}.
  \DOIprefix\doi{10.1371/journal.pone.0179473}.
\bibitem[{Wang et~al.(2017)Wang, Wu, Deng, Tang, Wang, Sun \&
  Shangguan}]{Wang2017}
\bibinfo{author}{Wang, Y.}, \bibinfo{author}{Wu, G.}, \bibinfo{author}{Deng,
  L.}, \bibinfo{author}{Tang, Z.}, \bibinfo{author}{Wang, K.},
  \bibinfo{author}{Sun, W.}, \& \bibinfo{author}{Shangguan, Z.}
  (\bibinfo{year}{2017}).
\newblock \bibinfo{title}{Prediction of aboveground grassland biomass on the
  loess plateau, china, using a random forest algorithm}.
\newblock {\it \bibinfo{journal}{Sci. Rep.}\/},  {\it \bibinfo{volume}{7}\/},
  \bibinfo{pages}{6940}. \URLprefix
  \url{https://doi.org/10.1038/s41598-017-07197-6}.
\bibitem[{Wenger \& Olden(2012)}]{Wenger2012}
\bibinfo{author}{Wenger, S.~J.}, \& \bibinfo{author}{Olden, J.~D.}
  (\bibinfo{year}{2012}).
\newblock \bibinfo{title}{Assessing transferability of ecological models: an
  underappreciated aspect of statistical validation}.
\newblock {\it \bibinfo{journal}{Methods in Ecology and Evolution}\/},  {\it
  \bibinfo{volume}{3}\/}, \bibinfo{pages}{260--267}. \URLprefix
  \url{https://besjournals.onlinelibrary.wiley.com/doi/abs/10.1111/j.2041-210X.2011.00170.x}.
  \DOIprefix\doi{10.1111/j.2041-210X.2011.00170.x}.
  \href{http://arxiv.org/abs/https://besjournals.onlinelibrary.wiley.com/doi/pdf/10.1111/j.2041-210X.2011.00170.x}{\tt
  arXiv:https://besjournals.onlinelibrary.wiley.com/doi/pdf/10.1111/j.2041-210X.2011.00170.x}.
\bibitem[{Xie et~al.(2017)Xie, Eftelioglu, Ali, Tang, Li, Doshi \&
  Shekhar}]{Xie2017}
\bibinfo{author}{Xie, Y.}, \bibinfo{author}{Eftelioglu, E.},
  \bibinfo{author}{Ali, R.~Y.}, \bibinfo{author}{Tang, X.},
  \bibinfo{author}{Li, Y.}, \bibinfo{author}{Doshi, R.}, \&
  \bibinfo{author}{Shekhar, S.} (\bibinfo{year}{2017}).
\newblock \bibinfo{title}{Transdisciplinary foundations of geospatial data
  science}.
\newblock {\it \bibinfo{journal}{ISPRS International Journal of
  Geo-Information}\/},  {\it \bibinfo{volume}{6}\/}. \URLprefix
  \url{http://www.mdpi.com/2220-9964/6/12/395}.
  \DOIprefix\doi{10.3390/ijgi6120395}.
\bibitem[{Yang et~al.(2016)Yang, Zhang, Yang, Zhi, Yang, Liu, Zhao \&
  Li}]{Yang2016}
\bibinfo{author}{Yang, R.-M.}, \bibinfo{author}{Zhang, G.-L.},
  \bibinfo{author}{Yang, F.}, \bibinfo{author}{Zhi, J.-J.},
  \bibinfo{author}{Yang, F.}, \bibinfo{author}{Liu, F.}, \bibinfo{author}{Zhao,
  Y.-G.}, \& \bibinfo{author}{Li, D.-C.} (\bibinfo{year}{2016}).
\newblock \bibinfo{title}{Precise estimation of soil organic carbon stocks in
  the northeast tibetan plateau}.
\newblock {\it \bibinfo{journal}{Sci. Rep.}\/},  {\it \bibinfo{volume}{6}\/},
  \bibinfo{pages}{21842}. \URLprefix \url{http://dx.doi.org/10.1038/srep21842}.
\bibitem[{Zanella et~al.(2017)Zanella, Folkard, Blackburn \&
  Carvalho}]{Zanella2017}
\bibinfo{author}{Zanella, L.}, \bibinfo{author}{Folkard, A.~M.},
  \bibinfo{author}{Blackburn, G.~A.}, \& \bibinfo{author}{Carvalho, L. M.~T.}
  (\bibinfo{year}{2017}).
\newblock \bibinfo{title}{How well does random forest analysis model
  deforestation and forest fragmentation in the brazilian atlantic forest?}
\newblock {\it \bibinfo{journal}{Environmental and Ecological Statistics}\/},
  {\it \bibinfo{volume}{24}\/}, \bibinfo{pages}{529--549}. \URLprefix
  \url{https://doi.org/10.1007/s10651-017-0389-8}.
  \DOIprefix\doi{10.1007/s10651-017-0389-8}.

\end{thebibliography}
\end{document}